%% file: JASA-template.tex
\documentclass[12pt]{article}
\usepackage{amsmath}
\usepackage{graphicx,psfrag,epsf}
\usepackage{enumerate}
\usepackage{natbib}
\usepackage{url} 

\usepackage{bm}
\usepackage{amssymb}
\usepackage{dsfont}
\usepackage{thmtools} 
\usepackage{thm-restate}
\usepackage{booktabs}
\usepackage{adjustbox}
\usepackage{xcolor}
\usepackage{hyperref}
\usepackage{setspace}
\usepackage{multicol}
\usepackage{subcaption}
\usepackage{mwe}

\usepackage{titlesec}

\usepackage[figuresright]{rotating}
\usepackage{authblk}

\newcommand{\blind}{1}

\begin{document}

\def\spacingset#1{\renewcommand{\baselinestretch}%
{#1}\small\normalsize} \spacingset{1}

\if1\blind
{
  \title{\bf Mixed Effects Spectral Vector Autoregressive Model: With Application to Brain Connectivity}
  \author{Anastasiia Malinovskaia\\
    Department of Statistics,  King Abdullah University of Science and Technology (KAUST)}
  \maketitle
} \fi

\if0\blind
{
  \begin{center}
    {\LARGE\bf Mixed Effects Spectral Vector Autoregressive Model: With Applications to Brain Connectivity}

\end{center}
} \fi


\begin{abstract}
The primary goal of this paper is to develop a method that 
quantifies how activity in one brain region can explain future activity in another region. Here, we propose the mixed effects spectral vector-autoregressive (ME-SpecVar) model to investigate differences in dynamics of dependence in a brain network between healthy children and those who diagnosed with ADHD. Specifically, ME-SpecVar model will be used to formally test for significant connectivity structure obtained using filtered EEG signals in delta, theta, alpha, beta, and gamma frequency bands. Suggested model allows one stage procedure for deriving Granger causality in common group structure and variation of subject specific random effects in different frequency oscillations. The model revealed novel results and showed more significant connections in all frequency bands and especially in slow waves in control group. In contrast, children with ADHD shared a pattern of diminished connectivity and variability of random effects. The results are consistent with previous findings about decreased anterior-posterior connectivity in children with ADHD. Moreover, the novel finding is that most diverse subject specific effective connectivity parameters in healthy children belong to parietal-occipital region that is associated with conscious visual attention.
\end{abstract}

\noindent%
Vector autoregressive model;
\vfill

\newpage
\spacingset{1.45} 

\input{1_intro}

\input{2_method}
\input{3_application}

\input{4_conclusion}

\clearpage
\newpage
\bibliographystyle{agsm}
\bibliography{5_references}

\end{document}

%% file: 1_intro.tex
\section{Introduction}\label{chap:intro}

\noindent 
\subsection{Effective connectivity}

The goal in this paper is to develop a statistical model and tool that will be used to investigate differences in brain effective connectivity between children diagnosed with attention deficit and hyperactivity disorder (ADHD) and healthy controls. There are several challenges to this problem. First, the measure of brain effective connectivity needs to go beyond the usual cross-correlation which captures some form of dependence but is inadequate for characterizing brain networks. Here, we will explore other meaningful and interpretable measures of effective connectivity (EC). The second challenge is that brain responses vary between people within groups and between groups. Thus, we have to develop a statistical model that accounts for both between-subject variation and between-group differences of EC. Lastly, it is important to study cross-relationships within different oscillatory activity of brain signals. Here, a novel statistical model was developed primarily for analyzing connectivity using electroencephalograph (EEG) signals collected from the scalp. Electroencephalography (EEG) and functional magnetic resonance imaging (fMRI) are the main non-invasive methods for collecting information reflecting underlying brain functions. Based on the type of the data and research question, different models for effective connectivity can be chosen. \par
Measures of effective connectivity can be roughly classified into three classes: Granger causality       \citep{GRANGER1980329}, information-theoretic measures \citep{PhysRevLett.85.461}  and dynamic causal modelling \citep{FRISTON20031273}.  Dynamic causal modelling is not an exploratory model because it assumes certain a-priori input parameters which are used for specific hypothesis testing. It captures non-linear interactions between brain regions, however the network size is limited \citep{FRISTON20031273}. Transfer entropy (TE) is model-free example of information flow-based theories. Kullback–Leibler divergence is a key moment in the model and it treats conditional probabilities of coupled regions \citep{https://doi.org/10.1002/hbm.20382}. Using TE, the analysis is applied without prior regions selection on whole-brain electrodes and the implementation is robust of model misspecification \citep{doi:10.1142/S0129065709001951}, however it can require sufficiently large number of time points. Other conventional way of characterizing effective connectivity is Granger Causality based on vector autoregressive models. Partial directed coherence \citep{SAMESHIMA199993} and directed transfer function \citep{KORZENIEWSKA2003195} are frequency domain extensions and hence they give distinct characterizations of EC across various frequency bands. One major limitation of this approach is that it captures only linear interactions, meanwhile EEG signals contain non-linear dependencies \citep{LopesdaSilva1989}. In addition, whole brain electrodes implementation of VAR models is problematic due to the curse of dimensionality and overparameterization. Implementing the VAR model with $r$ channels will have $O(r^2)$ number of parameters which can be computationally costly. Thus, penalized regression techniques and dimension reduction can be used to alleviate the problem \citep{https://doi.org/10.1111/j.2517-6161.1996.tb02080.x, Hu2019}.

The models of effective connectivity are relevant and give insights about cognition. Most of the cognitive functions are anatomically distributed across functionally specialized brain regions. Mechanisms that coordinate neuronal assemblies are important for understanding normal cognitive functioning and neuropathological conditions and diseases. The main novel contribution in this paper is a model that accounts for effective connectivity at different frequency bands to gain new insights about lead-lag dynamics and interconnections of brain regions at corresponding frequencies. Results obtained from the model might be linked to existing knowledge about particular frequency oscillations correlations with cognitive functions and abnormalities.

\subsection{Effective connectivity in ADHD}Non-invasive collection of neuronal activity data is the main source of brain information nowadays. Electroencephalogram (EEG), magnetoencephalogram (MEG) and fMRI techniques are commonly used in neuroscience and clinical research. They reveal the aspects of abnormal brain activity and develop diagnostics. EEG based effective connectivity studies broadened understanding of epilepsy, Alzheimer disease, major depression disorder and attention deficiency hyperactivity disorder (ADHD) \citep{BLINOWSKA2017667,Saeedi2021,PARKER2018943}. We focus on ADHD findings in the paper and we conducted analysis for EEG collected from healthy and ADHD children provided by \cite{rzfh-zn36-20}. 

Results of effective connectivity studies on ADHD participants are based on different experimental conditions and methods, therefore they lack of consistent outcomes and concurrent patterns. In paper by Ueda and collegues (2020), children with ADHD had significant functional hyperconnectivity in the theta range during executive functional tasks but not during the resting-state (in the delta to beta range) compared with children without ADHD \citep{UEDA2020129}.
Another finding showed a resting state deficiency in connectivity in ADHD and overconnectivity within and between frontal hemispheres while stimulus presentation \citep{10.1093/cercor/bhl089}. Chabot and Serfontein (1996) stated reduced parietal and increased intrahemispheric coherence in frontal and central regions in children with ADHD \citep{CHABOT1996951}. The results of \cite{article} on functional connectivity  identified a significant difference between the two groups in the T5 and O2 electrodes in the delta and theta frequencies. In the paper \cite{Ekhlasi2021} posterior to anterior patterns of information flow in theta frequency bands were disrupted compared to healthy children. Information flow between anterior regions and connections from central and lateral parietal areas to Pz electrode were significantly higher in healthy individuals than in the ADHD group in the beta band. \par
In general, analysis showed abnormal hyperconnectivity for certain regions in specific frequency bands and overall lack of connectivity compared to healthy controls. Current approaches share several limitations. The main approaches do not account for subject specific variations within the group and an average is used to represent the group connectivity in the vast majority of papers. Another observation indicates the lack of work that can attribute brain effective connectivity on oscillatory activity. Our proposed ME-SpecVAR approach can overcome these limitations by using mixed effects model for filtered signals in different frequency bands. \par
\subsection{Overview of the proposed model: Mixed Effects Spectral VAR (ME-SpecVAR)}\par
Most statistical models of brain effective connectivity are able to make statements about differences between groups of patients (healthy vs disease) and between experimental conditions. A standard approach for such kind of questions is to find estimates for each subject individually, and then average parameters inside each condition or group. Although, these approaches can be ineffective because they do not account for between-subject variation. Moreover, in some paradigms it could be essential to account for variation across subjects.
Linear mixed effects vector autoregressive model (ME VAR) \citep{GORROSTIETA20123347} tackles the accounting uncertainty of subject specific estimates and the authors proposed method of single-stage estimation of random variation between subjects. Moreover, Granger causality and connectivity differences between experimental conditions are captured in the former study. In ME VAR model connectivity matrix is decomposed into condition specific fixed component and participant-specific random effect. Analogously to the aforementioned approach, the model and analysis in the current paper are modified for different problem statement and expanded for frequency domain analysis that reveals new insights about connectivity structure between groups and subject specific variability in ADHD and healthy children.

The transition from time domain to the frequency domain is quite natural in analyzing brain electrophysiological signals. Spectral decomposition using filtering allows to provide analysis on extracted components and to infer connectivity at specific bands. The transition to ($\delta, \theta, \alpha, \beta,\gamma$) frequency bands enrich the analysis and it gives new insights on the problem. The choice of frequency division is based on the findings that each frequency band is associated with cognitive functions and abnormalities \citep{Hejazi2019,HERRMANN2004347}. Thus, if we find abnormal connectivity in particular frequency, we can link the results with existing knowledge of brain functioning at these frequencies.

The remainder of this paper is organized as follows. We introduce the formal construction of ME-SpecVar statistical model, describe simulation study and experimental paradigm along with aspects of children ADHD dataset, and conclude the paper with model implementation details, results and discussion.

%% file: 2_method.tex
\section{Mixed Effects Spectral VAR  Model (ME-SpecVAR)} \label{chap:method}

\subsection{General setting}

\noindent 
In this section, we describe our method to estimate group specific Granger Causality parameters and subject specific random effects based on vector autoregressive model extension to mixed effects framework \citep{GORROSTIETA20123347}. 

Suppose we have collected electroencephalogram signals from $R$ channels located on the scalp. Time-series vector $\mathbf{Y}^{(i)}({t})$  from participant $(i)$ represents data from $R$ channels:
\noindent
$\mathbf{Y}^{(i)}({t}) =\left[Y_1^{(i)}(t), Y_2^{(i)}(t),..., Y_R^{(i)}(t)\right], \text{where R is the number of EEG channels.}$
To capture the dependence structure between different channels a general additive model is proposed:
\begin{align}
\label{gen}
  \mathbf{Y}^{(i)}({t})=\mathbf{F}^{(i)}({t})+\mathbf{E}^{(i)}(t),  t=1, \ldots, T,  
\end{align}
where $\mathbf{F}^{(i)}({t})$ is the activity specific deterministic mean trend. After preprocessing procedures of EEG, $\mathbf{F}^{(i)}({t})$ is a zero mean process for each channel. If $\mathbf{F}^{(i)}({t}) = 0$, then subject specific random fluctuations $\mathbf{E}^{(i)}(t)=\mathbf{Y}^{(i)}({t})$ and  $$\mathbf{Y}^{(i)}(t) \sim \text{VAR(p)},$$ p - lag order of autoregression.

Hence, the model for preprocessed EEG signals is given by:
\begin{align}
\label{fluc}
 \mathbf{Y}^{(i)}(t)=\sum_{k=1}^{P}\left[\Phi_{1, k}^{(i)} G_{1}{(i)}+\Phi_{2, k}^{(i)} G_{2}{(i)}\right] \mathbf{Y}^{(i)}(t-k)+\mathbf{\epsilon}^{(i)}(t),
\end{align}
 where $\mathbf{\epsilon}^{(i)}(t)$ - zero mean white noise with covariance matrix that is common to all participants and here assumed to be diagonal, so that the cross-dependence structure will be  fully captured by the VAR coefficient matrices $\Phi_{1, k}^{(i)},\Phi_{2, k}^{(i)}$. The variable $G_1(i)$ is the indicator for the healthy control group ($1,...,n_1$) which takes the value of 1 if the $i-th$ subject belongs to this group. The variable $G_2(i)$  is the indicator for the ADHD group ($n_1+1,...,n_2$).
 

\noindent $\Phi_{1, k}^{(i)} \text{ and } \Phi_{2, k}^{(i)} -$ connectivity matrices for each participant (i), for each lag k, depending on the group (1 or 2).

It is noted that there is a high level of variation across subjects even in the same group. For this reason, it is important for the model to capture this variation which is accomplished by adding subject-specific random effect to fixed group component. In the model we decompose subject specific connectivity matrices into fixed group effect and random participant specific fluctuations:
\begin{align*}
    \Phi_{1, k}^{(i)}&=\Phi_{1, k}+{b}_{k}^{(i)}\\
    \notag\Phi_{2, k}^{(i)}&=\Phi_{2, k}+{b}_{k}^{(i)},
\end{align*}
where $\Phi_{1, k}$, $\Phi_{2, k}$ are the fixed connectivity components for groups 1 and 2; and ${b}_{k}^{(i)}$ are the participant-specific random effects for each lag $k$.
The model for random fluctuation can vary for number of groups, lag values and accounting for random part in connectivity structure.
In the proposed model we can identify the optimal lag structure of dependence between channels, discriminate connectivity structure - Granger Causality between groups and detect differences of random effects variance of participants in two groups. 

To determine effective connectivity, we conduct statistical inference on the connectivity parameter $\Phi_{g,l}(r,r')$, where $g$ - group, $l$ - lag values from $1\dots p$, $(r,r')$ - two arbitrary channels. If $\Phi_{g,l}(r,r')$ is significantly non-zero, then we state that channel $r'$ improves predictability of channel $r$ and there exists Granger causal connection between them.  For each pair of channels, Welch-Satterthwaite T-Test (corrected for multiple testing) was conducted to compare significant components among two groups. Another way to test for significance of the lagged channel activity in the model is to conduct likelihood ratio test (LRT) using nested models \citep{buse}. The ratio is constructed by the full model likelihood including the lagged variable and likelihood for the reduced model with the lagged variable removed from the model.

\subsection{Proposed ME-SpecVAR}\label{chap:spectral}

\begin{figure}[!htp]
	\centering
    \title{One channel EEG signal spectral decomposition}\par
     \vspace{4mm}
	\begin{tabular}{c}
		\includegraphics[scale=0.7]{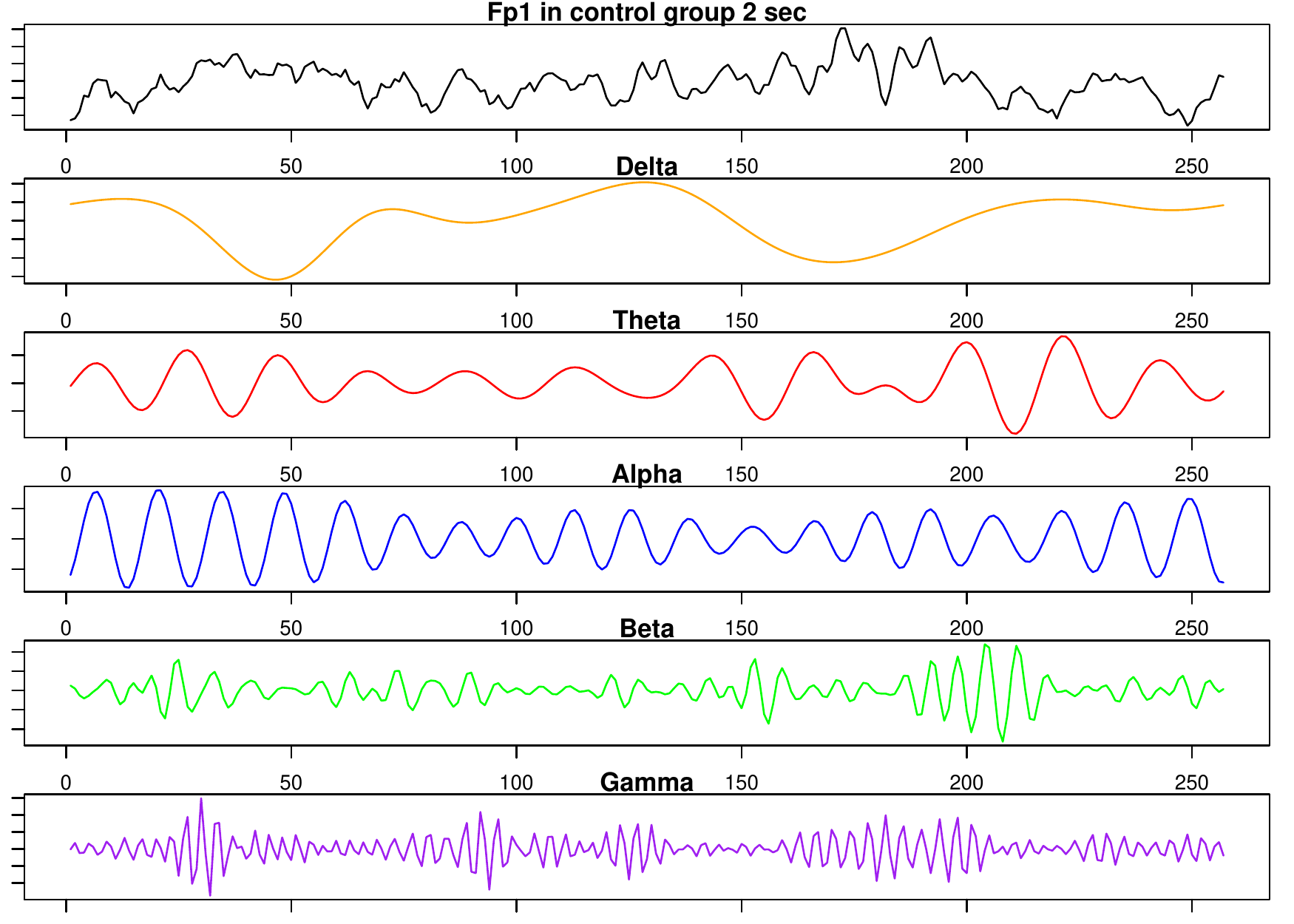}
	\end{tabular}
	\vspace{4mm}

	\caption{ Spectral decomposition into ($\delta, \theta, \alpha, \beta,\gamma$) frequency bands - filtered signals. Upper plot represents initial EEG signal from Fp1 channel with 2 seconds - 256 time-points. Spectral decomposition implemented using 3-rd order Butterworth filter.}
	\label{fig:spec}
\end{figure}

\noindent Electrical activity of neurons is essentially the summation of simultaneous neuronal firings, thus the resulting signal is a composition of underlying processes. The appearance of the signal reflects that the signal contains different periodic components that can be extracted as it is shown on \ref{fig:spec}.

Stationary time series can be presented as linear combination Fourier waveforms over different frequencies. The main idea is to apply a linear filter to EEG signals, so that obtained signal components have a power spectrum concentrated around a prespecified band. The choice of the bands is conventionally established by many neuroscientific studies. The commonly used ($\delta, \theta, \alpha, \beta,\gamma$) frequency components of brain EEGs are linked to specific consciousness states and cognitive functions, like sleep, active concentration and memory activation.  Alpha  frequency at 8-12 Hz is mostly inherent to closed eyes alert state \citep{doi:10.1177/0956797617699167}, beta (12 - 30 Hz) is associated with thinking and active concentration \citep{doi:10.1179/147683008X301478}, gamma (30 - 50 Hz) is mostly correlated with high order cognitive functions such as memory, attention. Slow waves as delta (0.5 - 4 Hz) and theta (4 - 8 Hz) are connected with sleep, learning and memory \citep{Etard5750}. During mental disorders abnormal activities of the rhythms are detected \citep{BASAR201319}. According to the meaningful results of the frequency domain analysis, we decided to combine using filtered signals and effective connectivity modeling.

For each participant $(i)$ and for each channel $r$, we can decompose the signals into commonly used frequency bands ($\delta, \theta, \alpha, \beta,\gamma$) using 3-rd order Butterworth filter \citep{ALARCON200035}.

After decomposing signals instead of vector of time series $\mathbf{Y}_r^{(i)}({t})$ we treat five new time series vectors $\mathbf{Y}_{r,band}^{(i)}({t}),$ where ${band} = (\delta, \theta, \alpha, \beta,\gamma).$ 
 Figure \ref{fig:spec}, (b.) represents the example of spectral decomposition of one channel data.
General additive model \eqref{gen} and model for random fluctuations \eqref{fluc} are included in ME-SpecVar and we will investigate effective connectivity among various oscillatory waveforms. Filtered components of the signals allow us to conduct time-series model estimation at specific frequency bands and obtain corresponding results.

One major contribution of this paper is a novel mixed effects approach to different frequency filtered EEG time series. The model parameters are obtained using one stage estimation of group specific connectivity and subject specific random effects. We conducted simulation study to verify the model adequacy and accuracy for simulated signals. Furthermore, we justify the choice of lag value in the model.

\subsection{Exploratory analysis}
\begin{figure}[!htp]
	\centering
    \title{LASSO VAR(4) estimation}\par
     \vspace{4mm}
	\begin{tabular}{c}
		\includegraphics[scale=0.4]{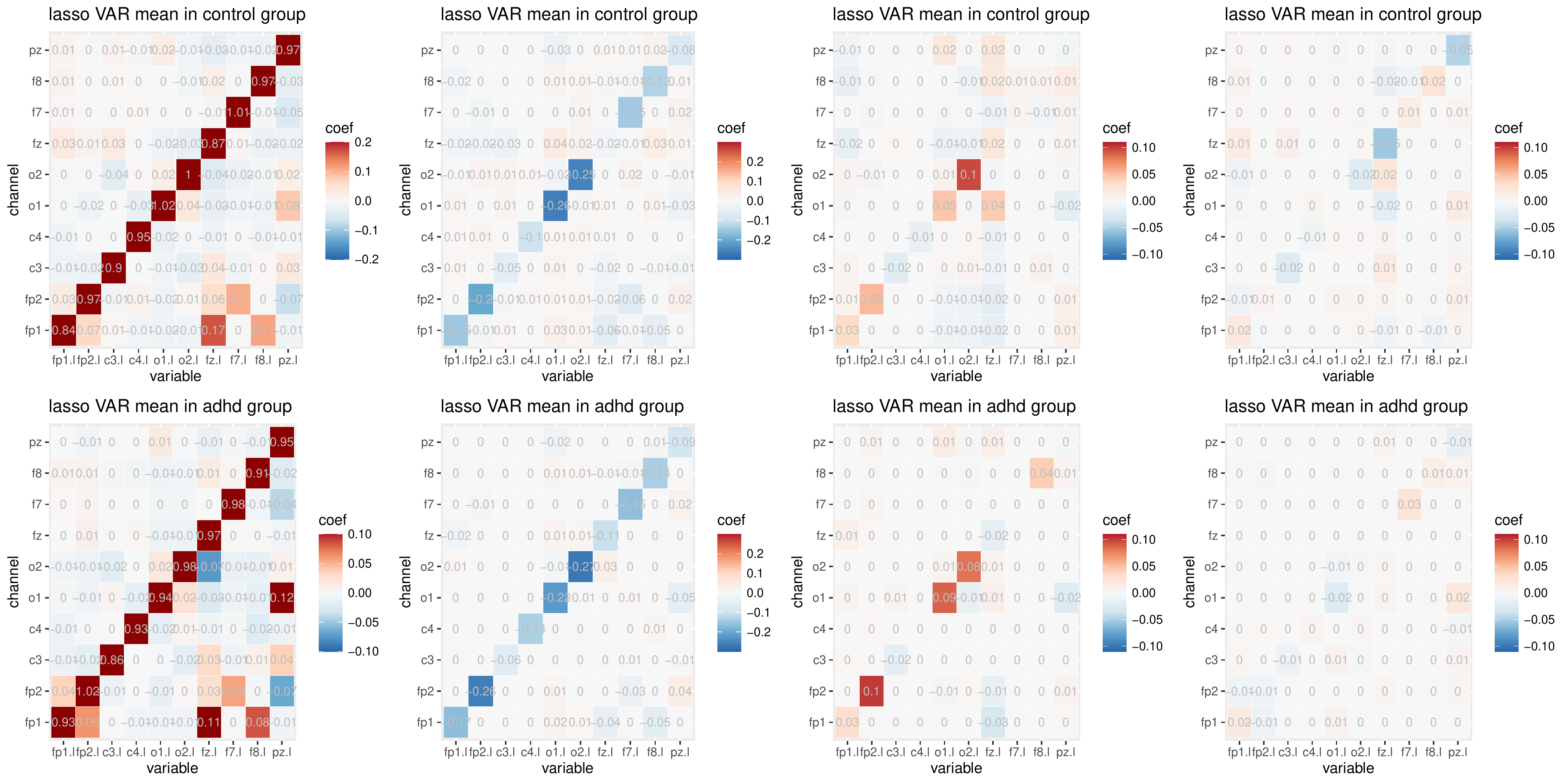}
	\end{tabular}
	\vspace{4mm}

	\caption{Mean values of the estimated LASSO+LSE VAR (4) parameters based on 10 channels, 30 seconds of data and 53 participants in control group and 51 participants in ADHD. The number of non-zero parameters is decreasing while lag value is increasing. }
	\label{fig:LASSO}
\end{figure}

\noindent Using exploratory data analysis, we applied the model selection algorithm for choosing the best lag value for the model based on LASSLE \citep{Hu2019} estimation procedure. For each participant we calculated the best lag value based on different information criteria, then according the obtained results Schwartz Criterion (SC or BIC) was the lowest and equal to 4. We investigated the parameters structure while lag value is increasing. The figure \ref{fig:LASSO} shows that the most of non-diagonal elements are close or equal to zero for lags more than 1. Whereas the general goal of the paper is to investigate cross dependence between EEG channels, we decided to proceed with the ME-SpecVAR(1) model containing the majority of between channels connections.

\subsection{Simulation study}
\begin{figure}[!htp]
	\centering

    \title{Bias, MSE, STD of fixed effects (a)}
    
    \vspace{4mm}
	\begin{tabular}{c}
	    \includegraphics[scale=0.43]{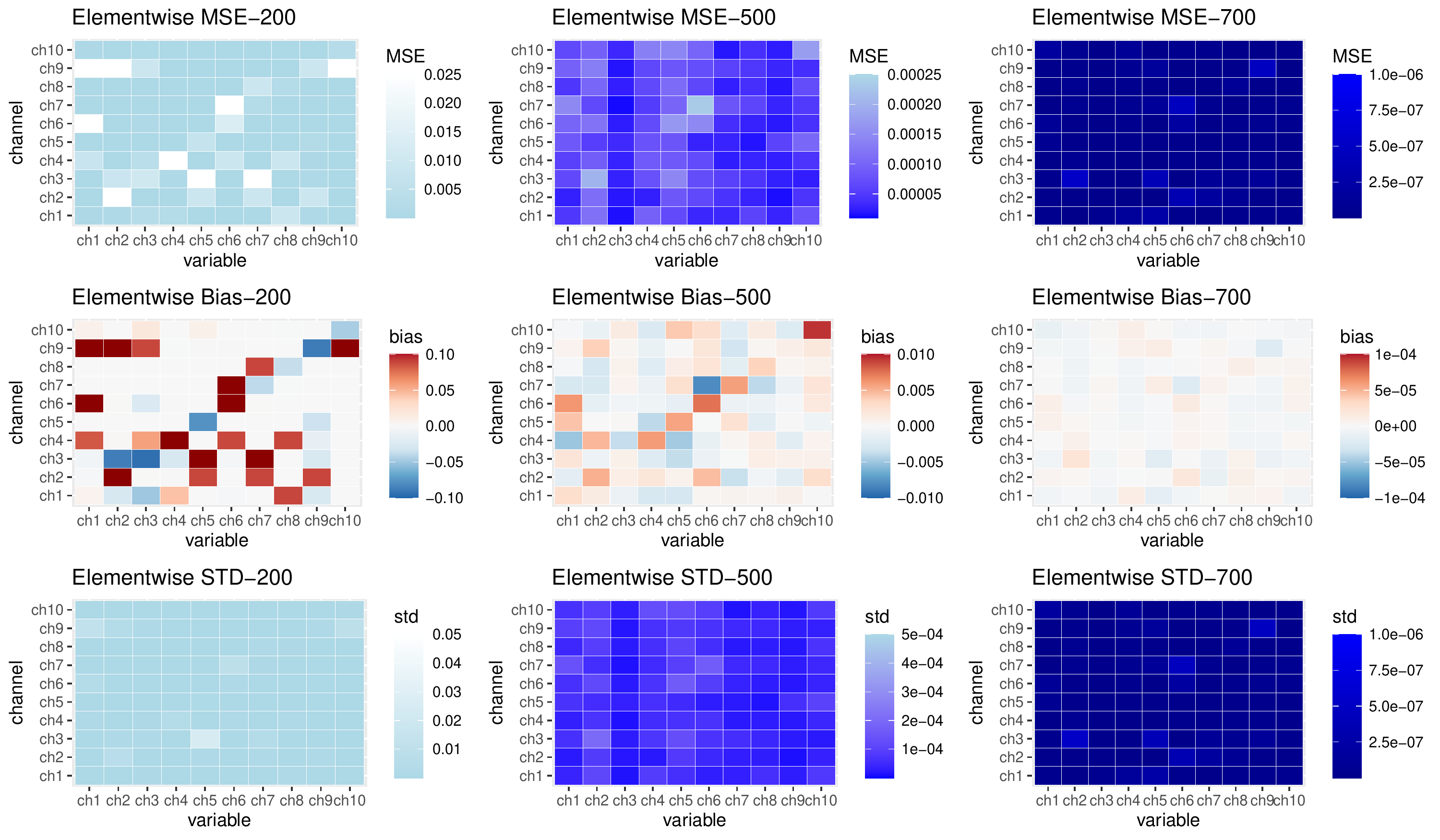}
	\end{tabular}
	\vspace{4mm}
	\centering
	
	\title{Bias, MSE, STD of random effects (b)}
	
	\vspace{4mm}
	\begin{tabular}{c}
		\includegraphics[scale=0.43]{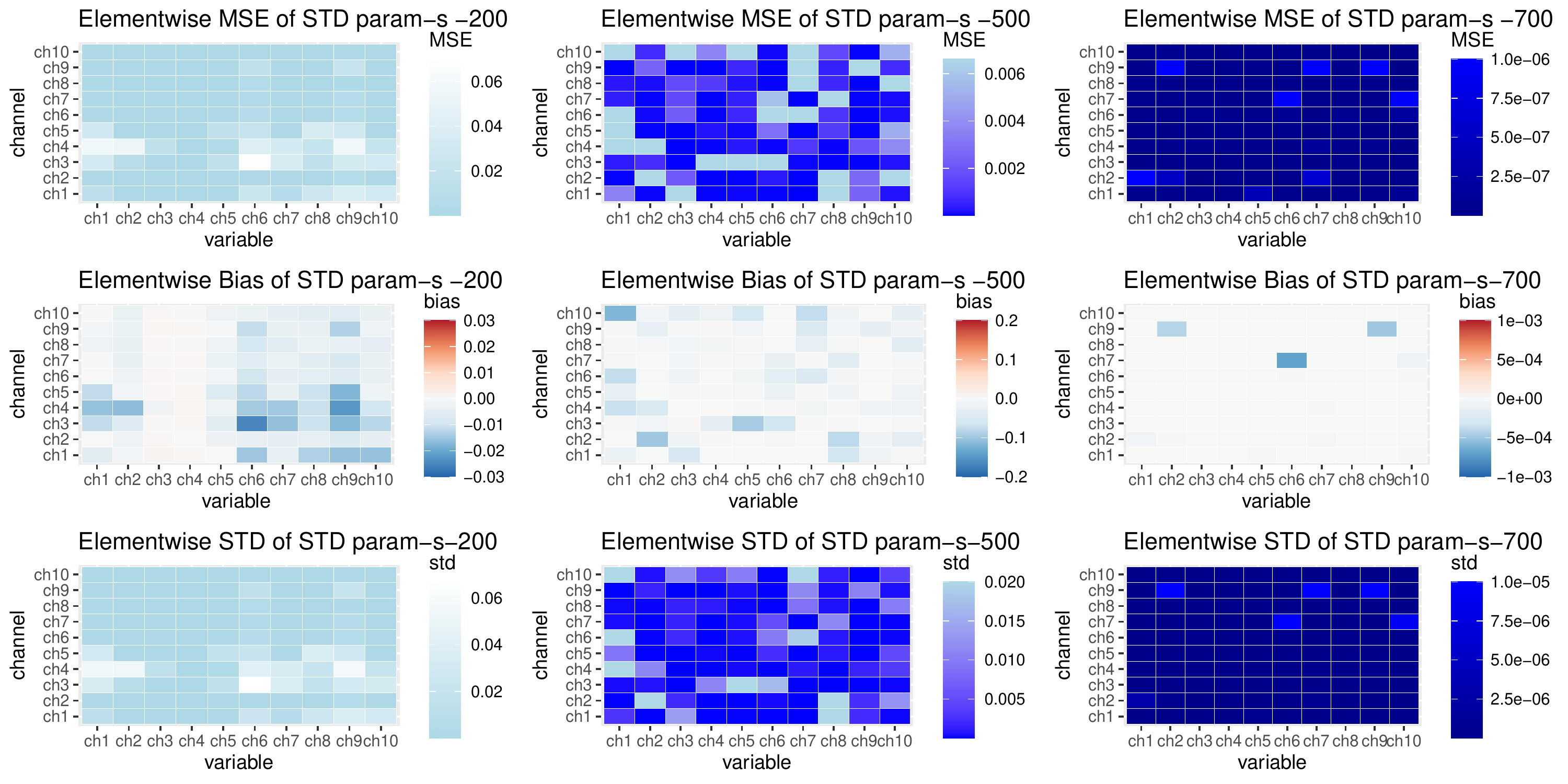}
	\end{tabular}
	
	\caption{Illustration of Mean Squared Error, Bias and  standard deviation of simulated vs ME-SpecVAR estimated data. As t - number of time points increases, MSE, Bias and STD decrease.}
	\label{fig:sim_study}
\end{figure}

We propose testing the model on generated time series data from VAR(2) model with additional random effects. The goal of the study is to assess how the model performs on realistic simulated data. The parameters of VAR(2) model consist of $\Phi_1$ and $\Phi_2$ matrices. Matrices of parameters were chosen to keep VAR(2) process causal. For each participant (i) 10 channels vector is $\mathbf{Y}^{(i)}({t}) =\left[Y_1^{(i)}(t), Y_2^{(i)}(t),..., Y_{10}^{(i)}(t)\right]$. $$\mathbf{Y}^{(i)}({t}) = \Phi_1^{(i)}\mathbf{Y}^{(i)}({t-1})+\Phi_2^{(i)}\mathbf{Y}^{(i)}({t-2})+ \mathbf{W^{(i)}}(t),$$  VAR(2) process for each participant (i), where $\Phi_1^{(i)}$ and  $\Phi_2^{(i)}$ are $10\times10 $ matrices,  $\mathbf{W^{(i)}}(t)$ - is white noise process with mean 0 and variance $\sigma_w^2$. Each participant matrices $\Phi_2^{(i)}$ and $\Phi_2^{(i)}$ are decomposed into common group specific component ($\Phi_1$ and $\Phi_2$) and random component. Random component is constructed in the following manner: it is a random matrix realization for each participant from normal distribution with zero mean and $\sigma_1^2$ and $\sigma_2^2$ variance for each lag parameters accordingly. Using ME-SpecVAR(1) estimation we will compare real group parameters ($\Phi_1$) and real standard deviations of random components ($\sigma_1^2$) for lag 1 with estimated. The motivation for restricting the model to lag 1 parameters instead of using both lag 1 and lag 2 is to check additionally the validity of the simplest lag 1 model in our real data application.  We simulated  10 channel realizations for 10 participants from the VAR(2) process for number of time points in 3 regimes: 200, 500, 700. We fit the ME-SpecVAR(1) model for each regime using each participant model realizations. The procedure was repeated for 10000 times in each regime. As a result, we have component-wise mean squared error, bias and standard deviation of common group $\Phi_1$ parameters and matrix of standard deviations of random effects. The results are illustrated in Figure \ref{fig:sim_study}.  
The results show that estimation of group specific parameters and random effects variations based on ME-SpecVAR(1) is mean-squared consistent. The error of estimation decreases with increasing time points regime. As we showed previously in exploratory analysis, most of the between channels significant connections are observed in lag 1 parameters. The simulation study justifies that the choice of simpler model is reasonable. Simpler model correctly captures connectivity between components at lag 1 and significantly decreases computation time.

%% file: 3_application.tex
\section{Effective connectivity in ADHD}\label{chap:application}

\subsection{The children ADHD dataset}

\begin{figure}[!htp]
	\centering
    \title{Standard 10-20 system}\par
    \vspace{4mm}
    
	\begin{tabular}{c}
	    \includegraphics[scale=0.25]{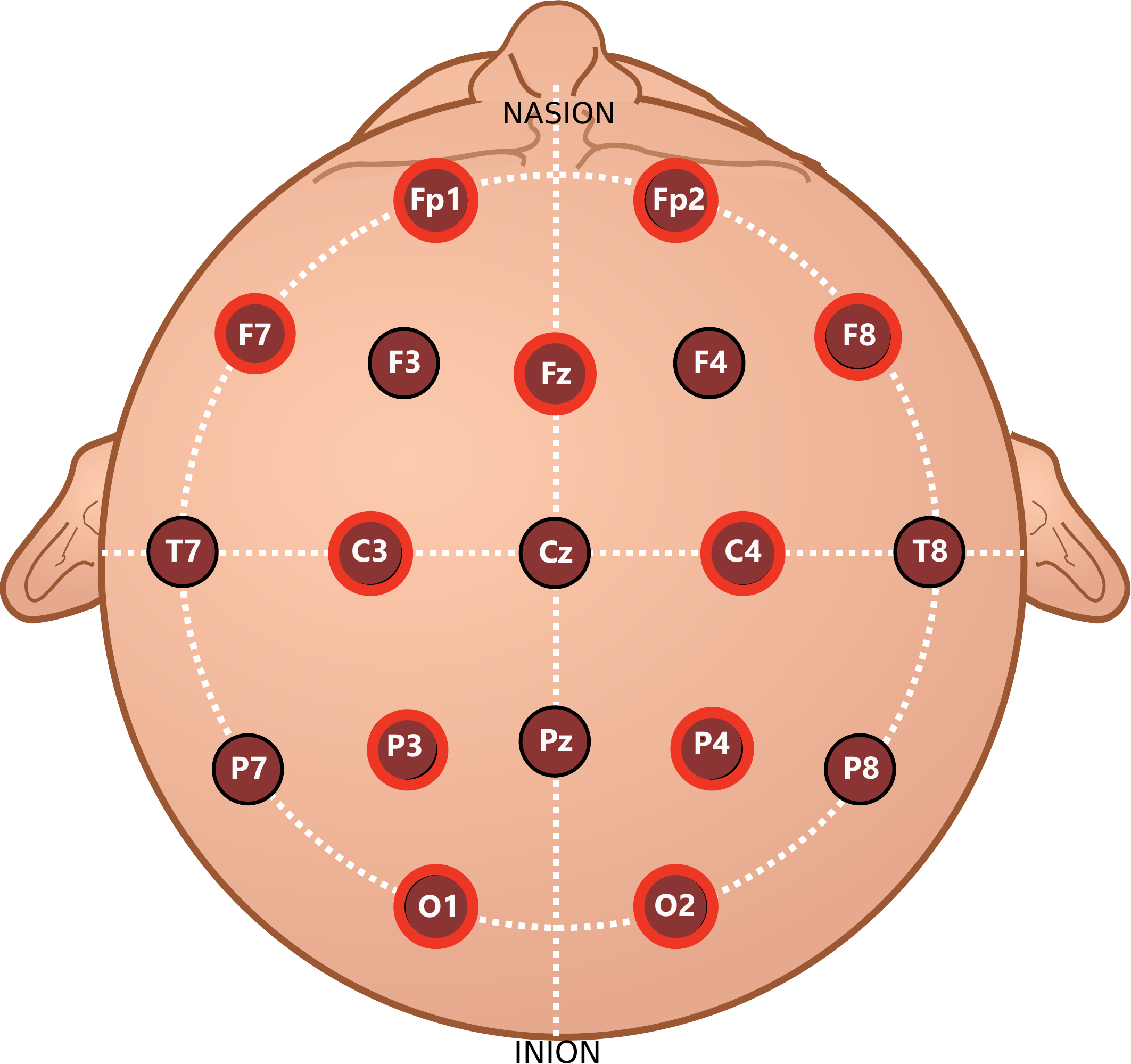}
	\end{tabular}

	\caption{Standard electrode position system 10-20 with highlighted channels for the analysis.}
	\label{fig:eeg}
\end{figure}

\noindent The data was collected by \cite{rzfh-zn36-20} from  51 children with ADHD and 53 healthy controls (boys and girls, ages 7-12). The ADHD children were diagnosed and medicated for up to 6 months. The EEG recording was done using 10-20 standard electrodes setting with 19 channels at 128 Hz sampling frequency. Figure \ref{fig:eeg} displays a standard layout of EEG channels in the 10-20 system.

The goal of the experiment was to study attention to a visual task, as ability to sustain attention is a key aspect in Attention Deficit Hyperactivity Disorder. In particular, the children were asked to count cartoon characters. The number of characters randomly changed between 5 and 16. Stimulus was continuously presented without any break after participant's response. PREP pipeline was used to standardize EEG signals. Pre-processing contained removing the effect of the electrical line and artifacts due to eye movements,  removing eye blinks and muscular movements, bad quality channels detection, filter non-relevant signal components and re-referencing the signal.  Butterworth band-pass filter for segmenting the signal into the main brain rhythms was applied.
For current analysis we used 30 seconds data for each of 11 (Fp1, Fp2,Fz, F7,F8, C3, C4, P3, P4, O1, O2) channels highlighted in Figure \ref{fig:eeg}. The signals were scaled and outliers replaced.

\subsection{ME-Spec-VAR model implementation}
\begin{figure}[!htp]
	\centering

	\centering
    \title{Granger causality in Control and ADHD groups (a)}\par
    \vspace{4mm}
    
	\begin{tabular}{ccccc}

		\includegraphics[scale=0.14]{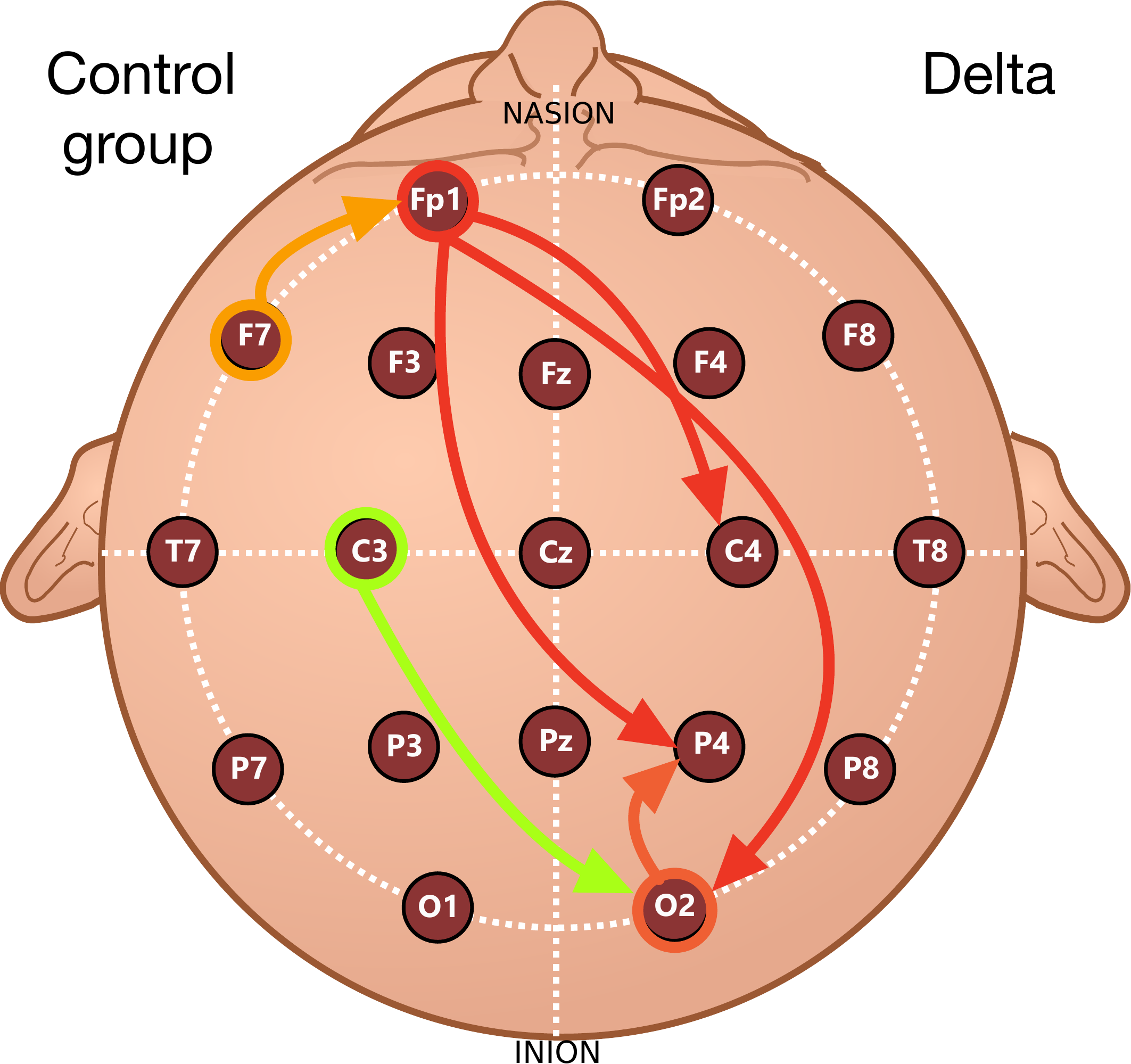}&
		\includegraphics[scale=0.14]{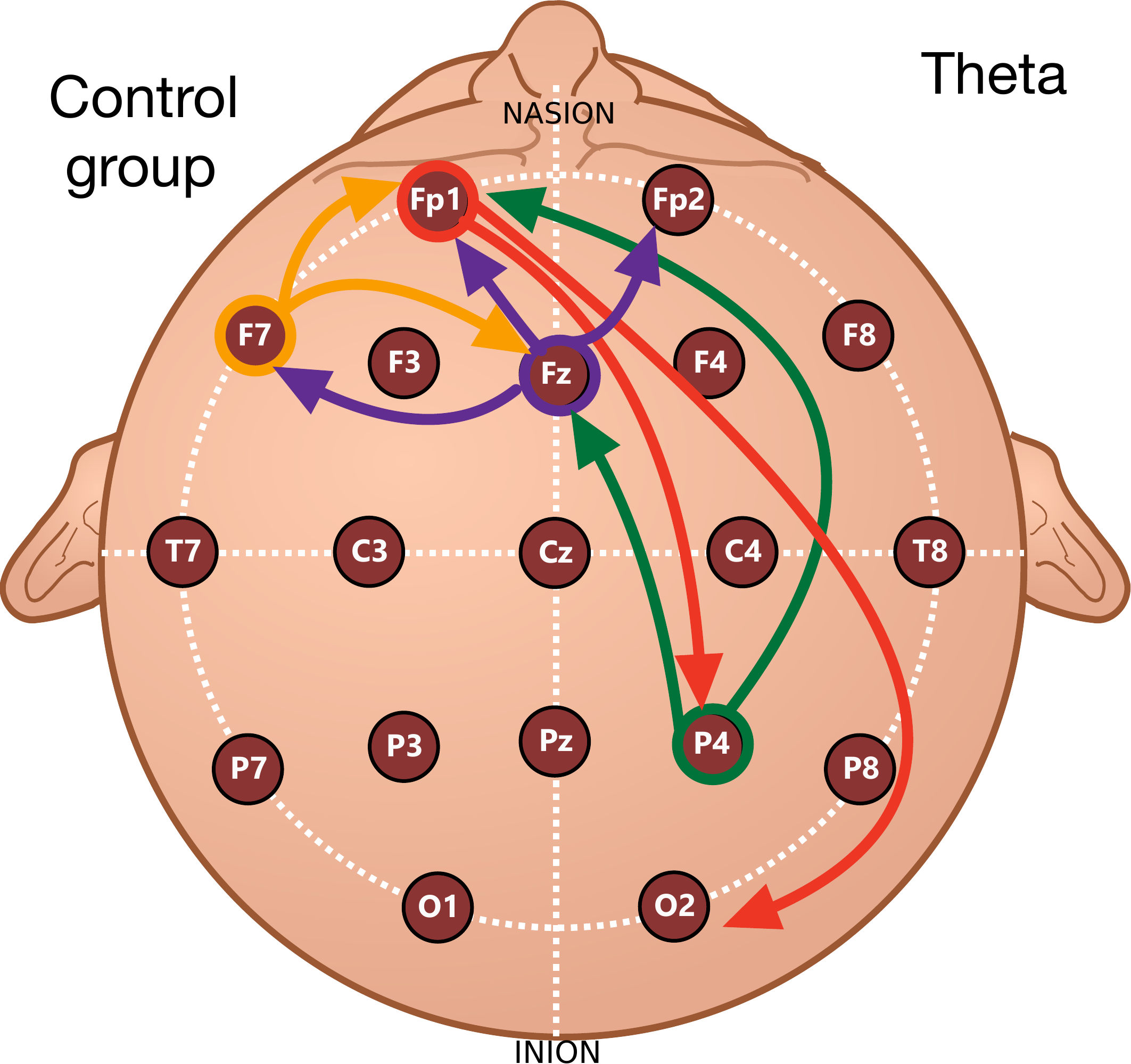}&
		\includegraphics[scale=0.14]{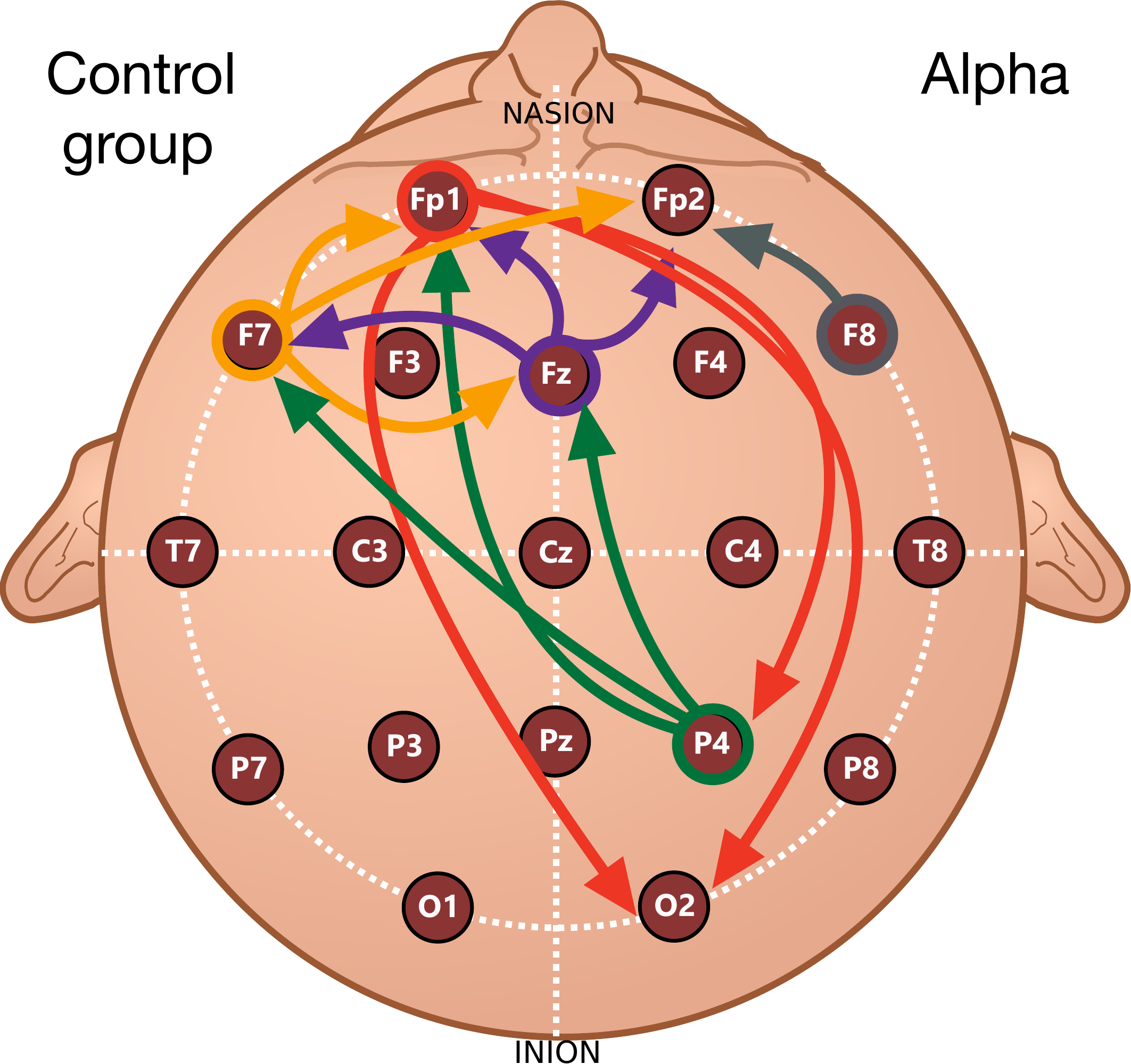}&
		\includegraphics[scale=0.14]{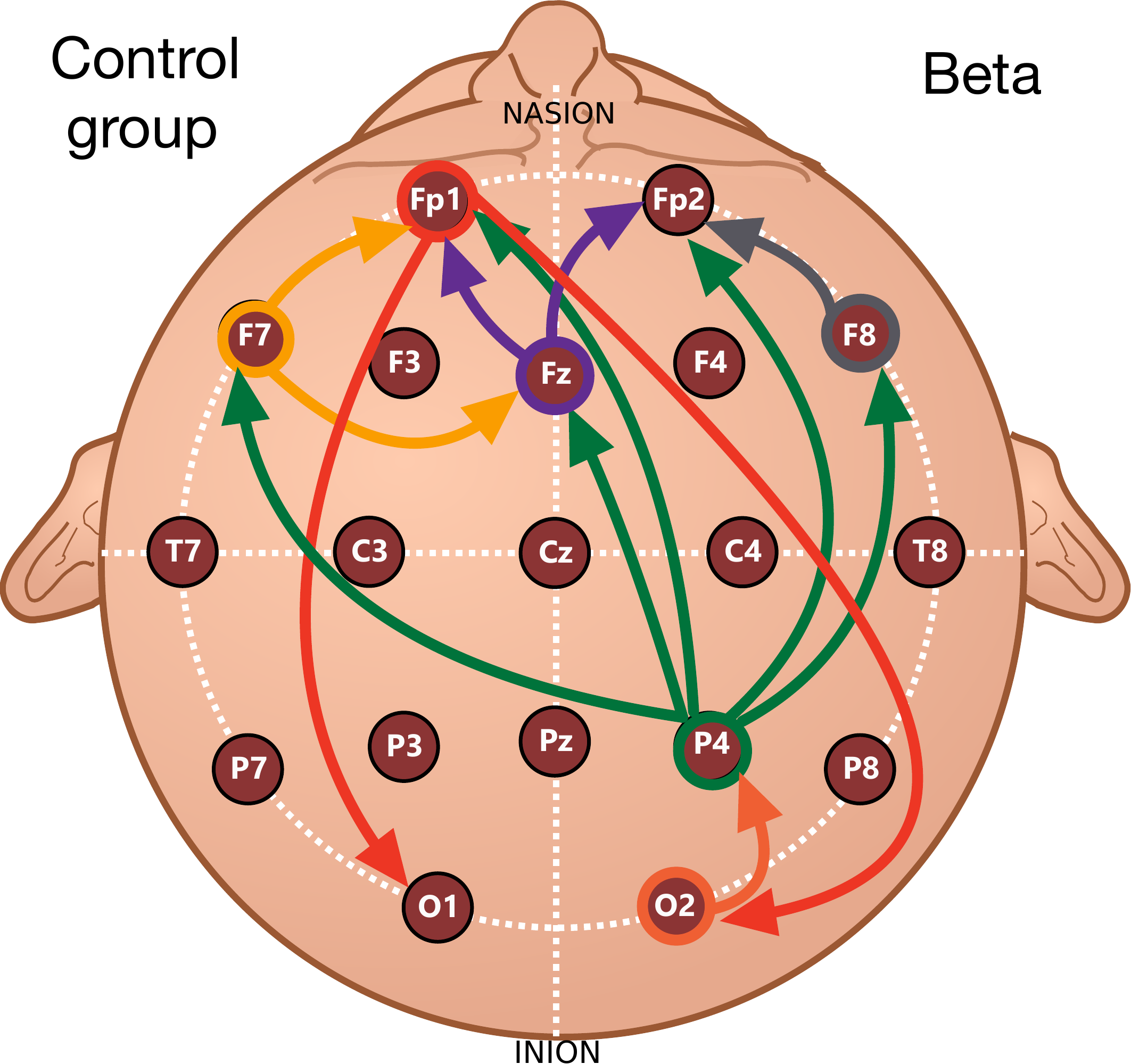}&
		\includegraphics[scale=0.14]{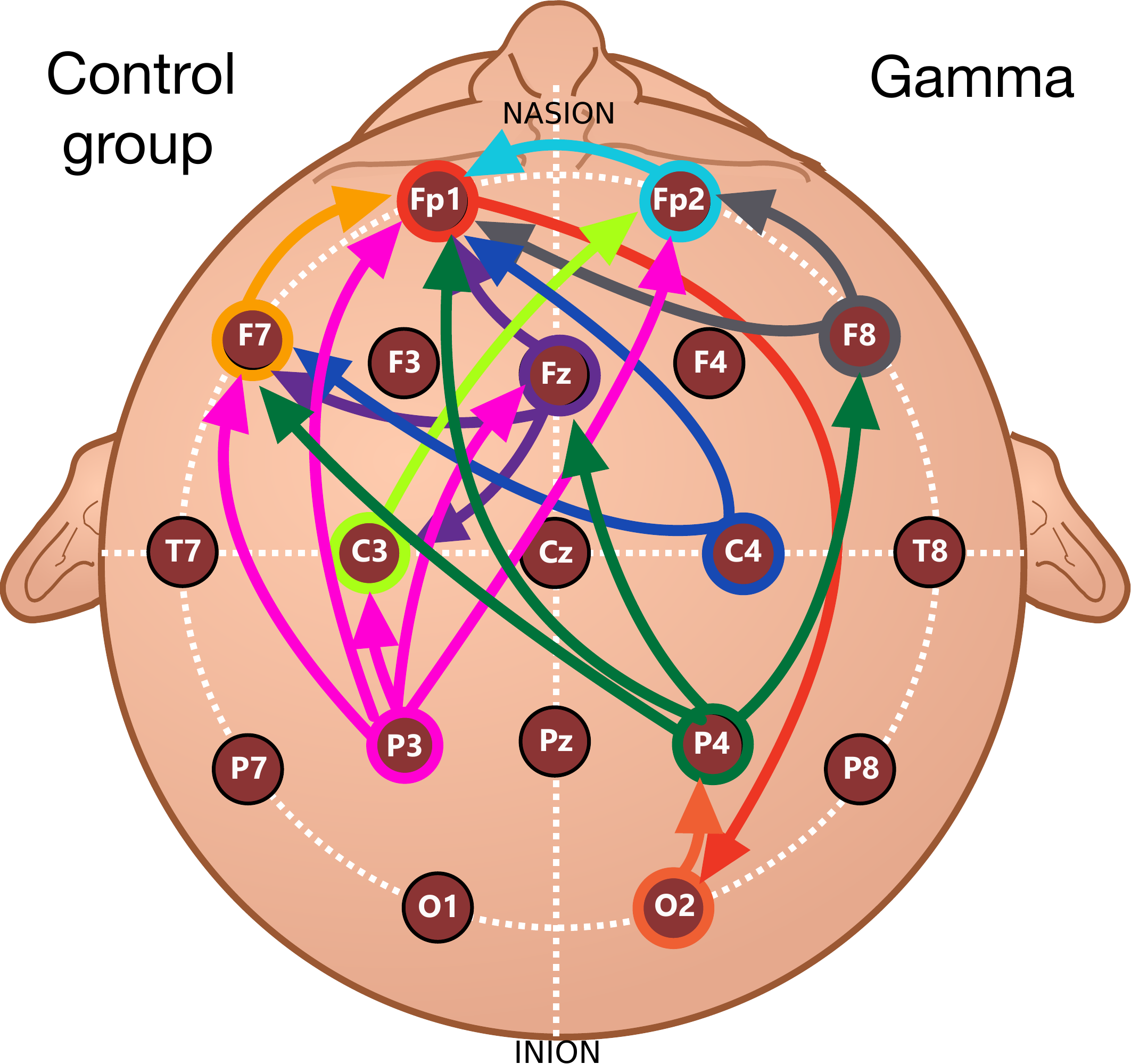}
	\end{tabular}
	\vspace{8mm}
	\begin{tabular}{ccccc}
		\includegraphics[scale=0.14]{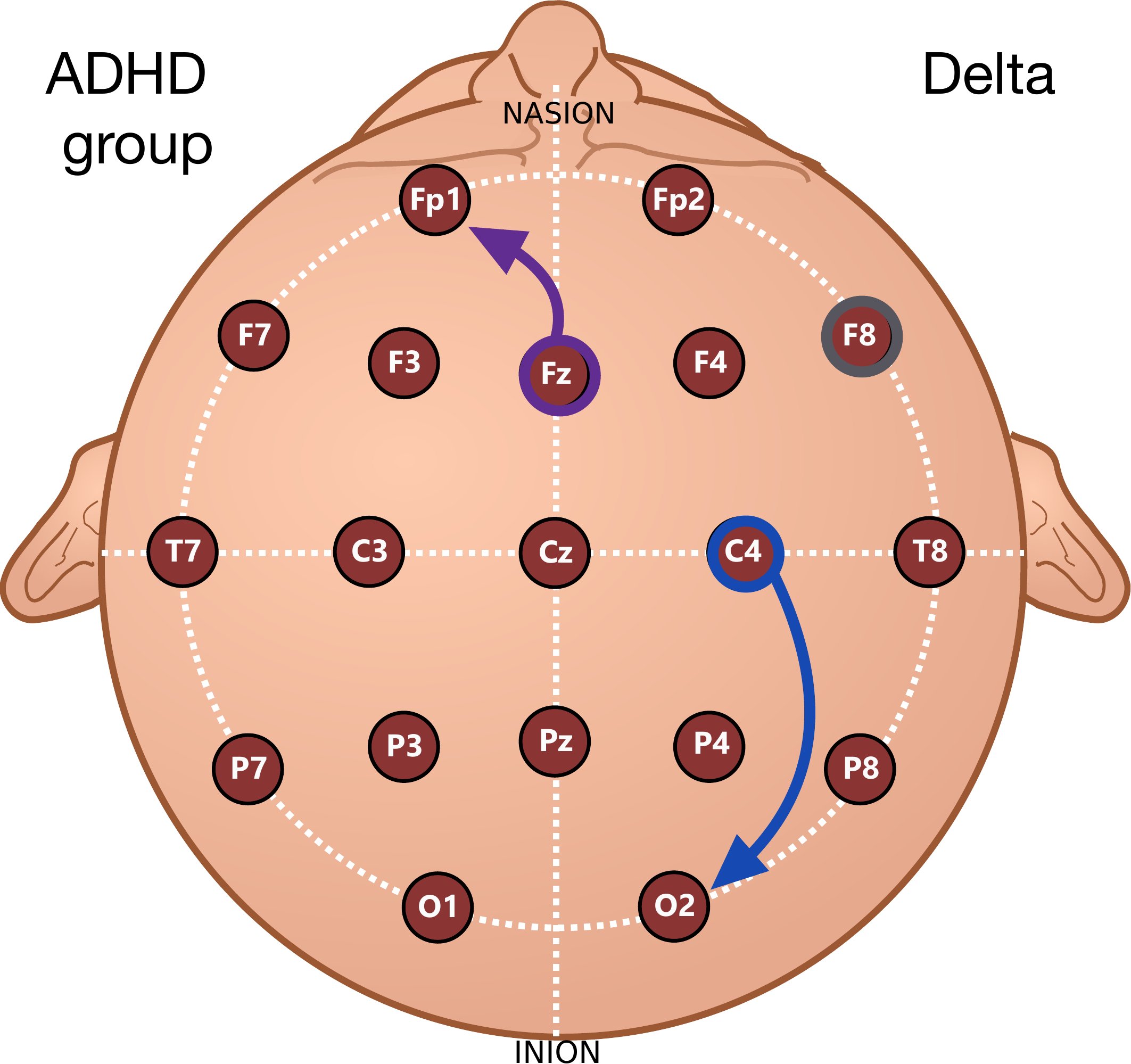}&
		\includegraphics[scale=0.14]{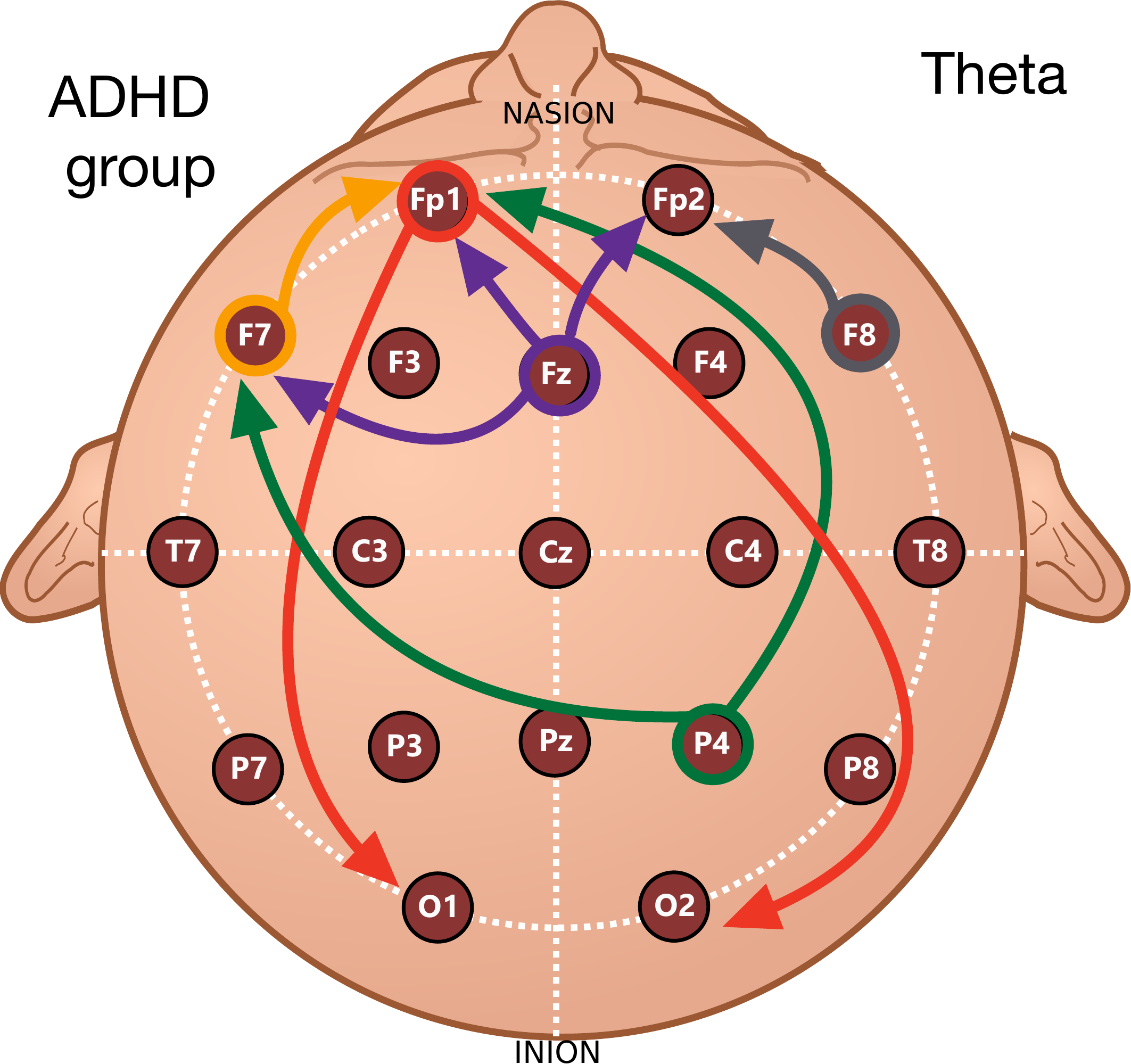}&
		\includegraphics[scale=0.14]{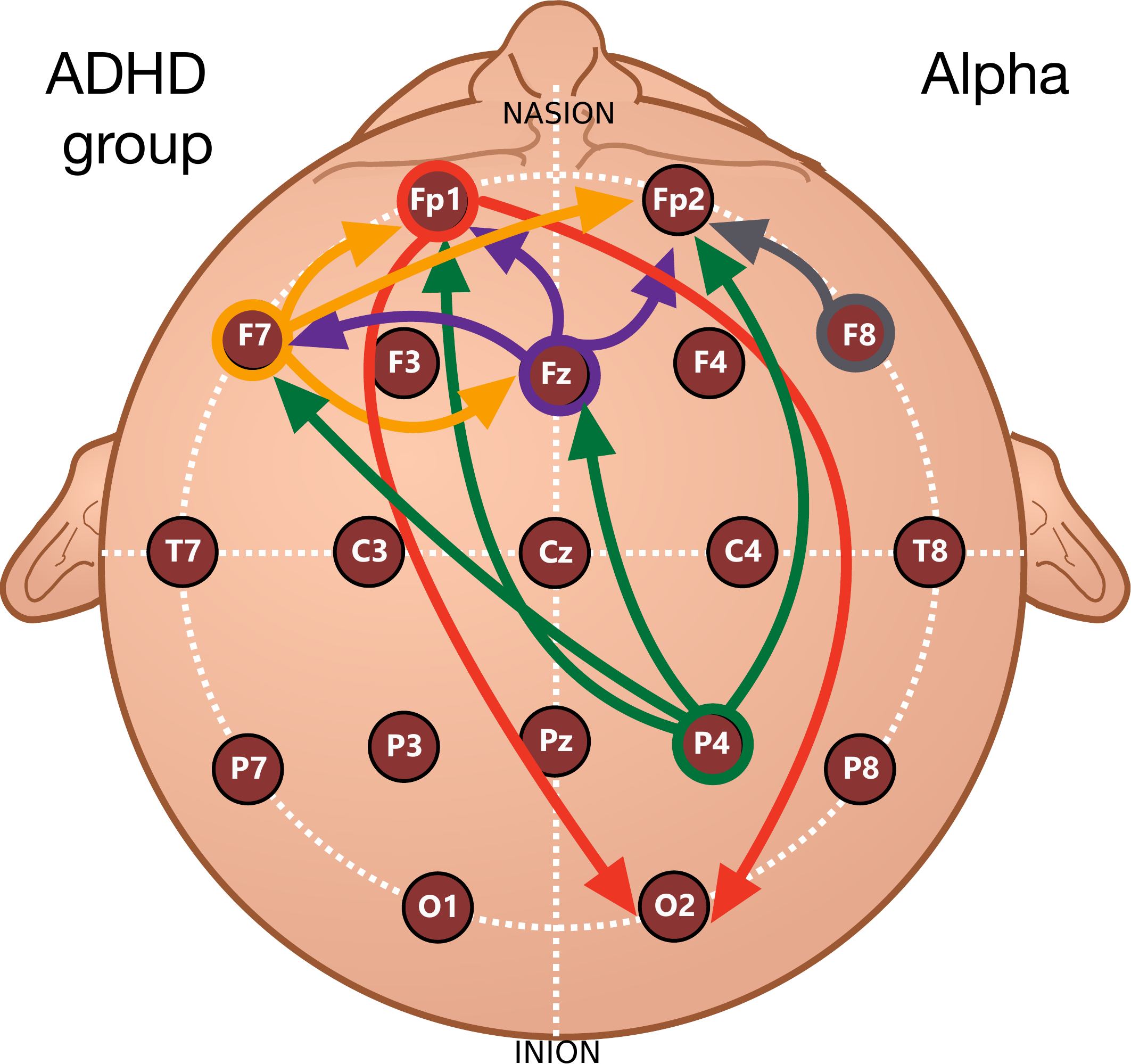}&
		\includegraphics[scale=0.14]{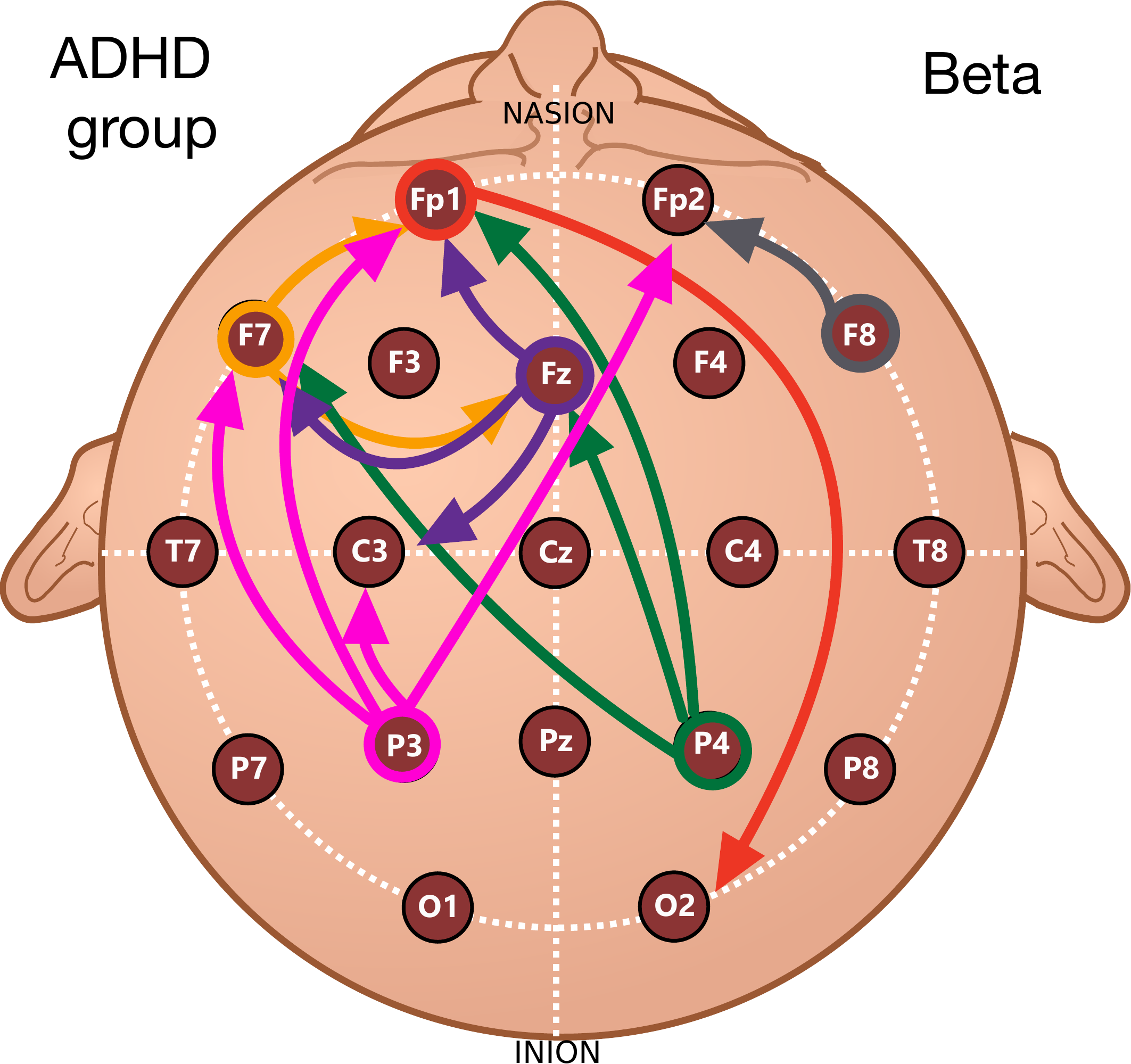}&
		\includegraphics[scale=0.14]{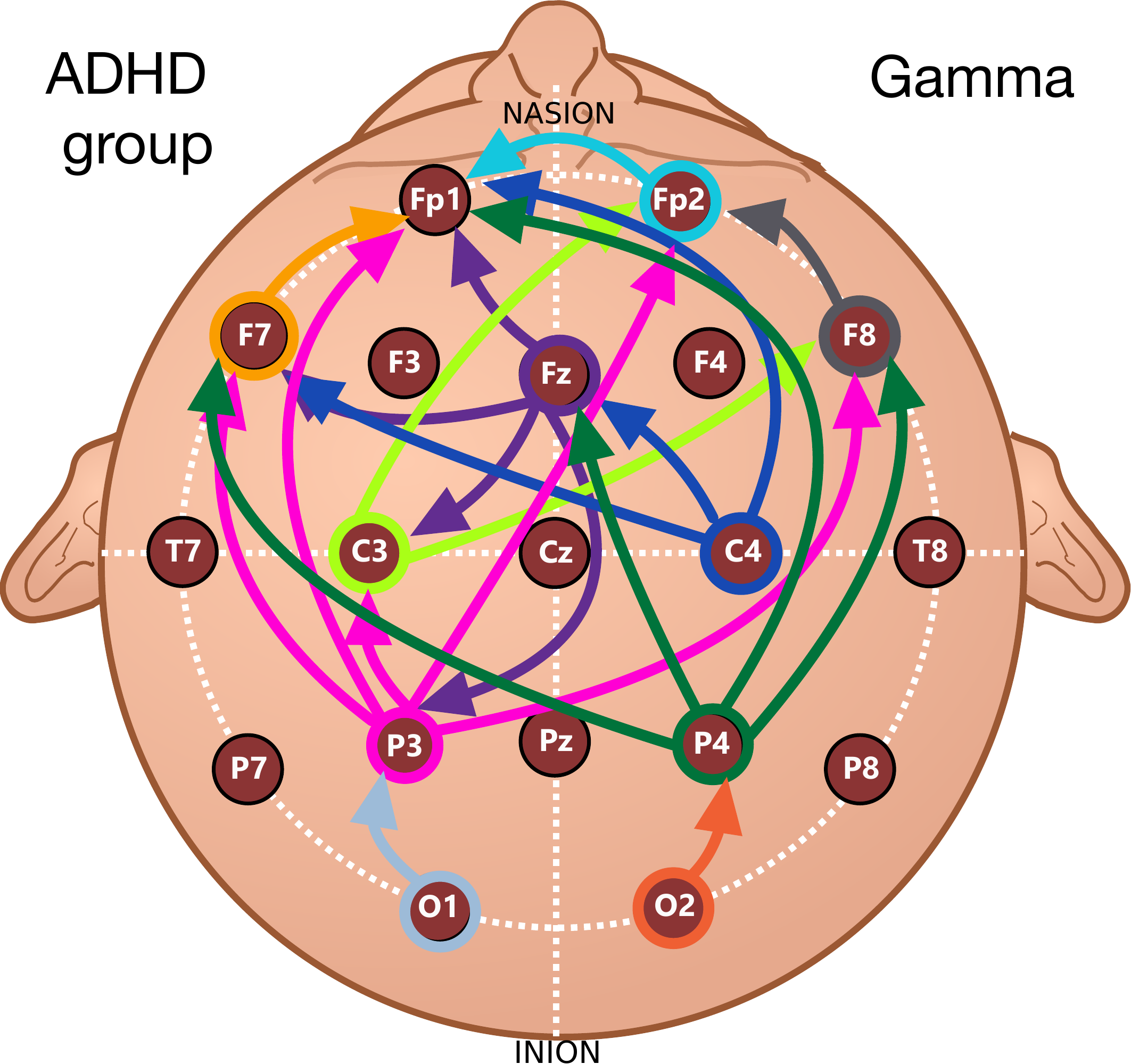}
	\end{tabular}
	\vspace{2mm}
	\vspace{2mm}
	\title{Differences in Granger Causality between groups (b)}\par

	\begin{tabular}{ccccc}
		\includegraphics[scale=0.14]{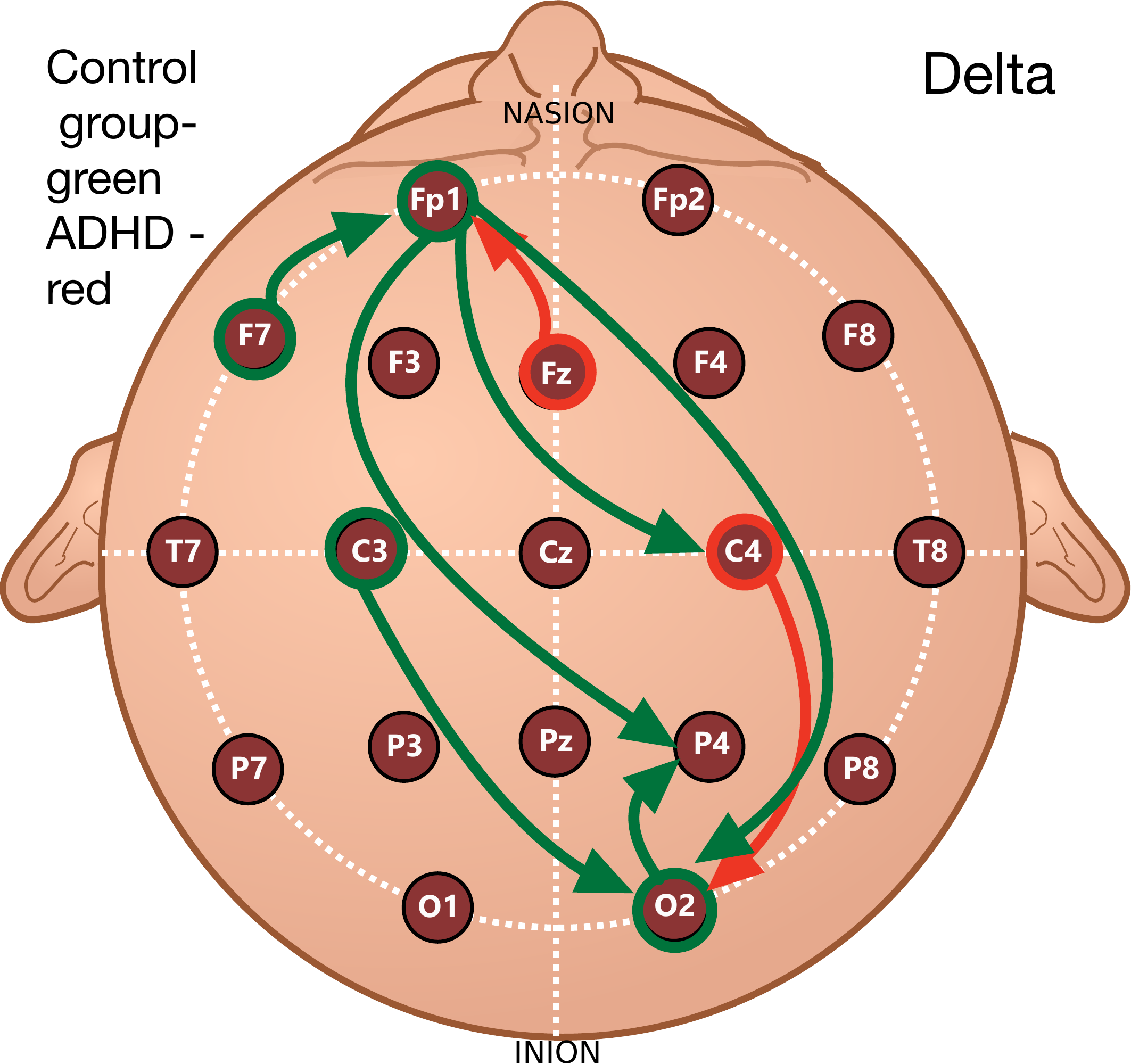}&
		\includegraphics[scale=0.14]{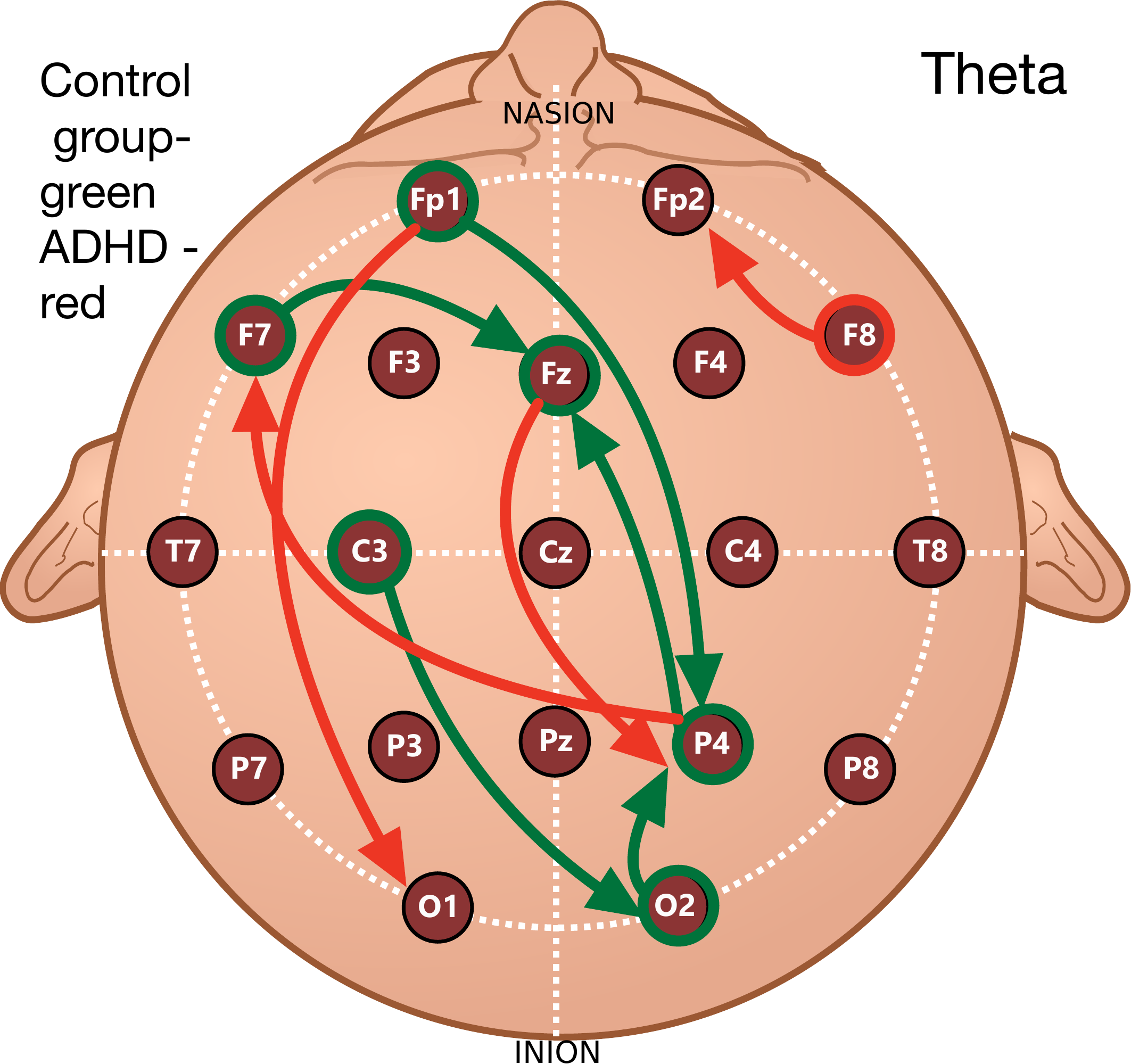}&
		\includegraphics[scale=0.14]{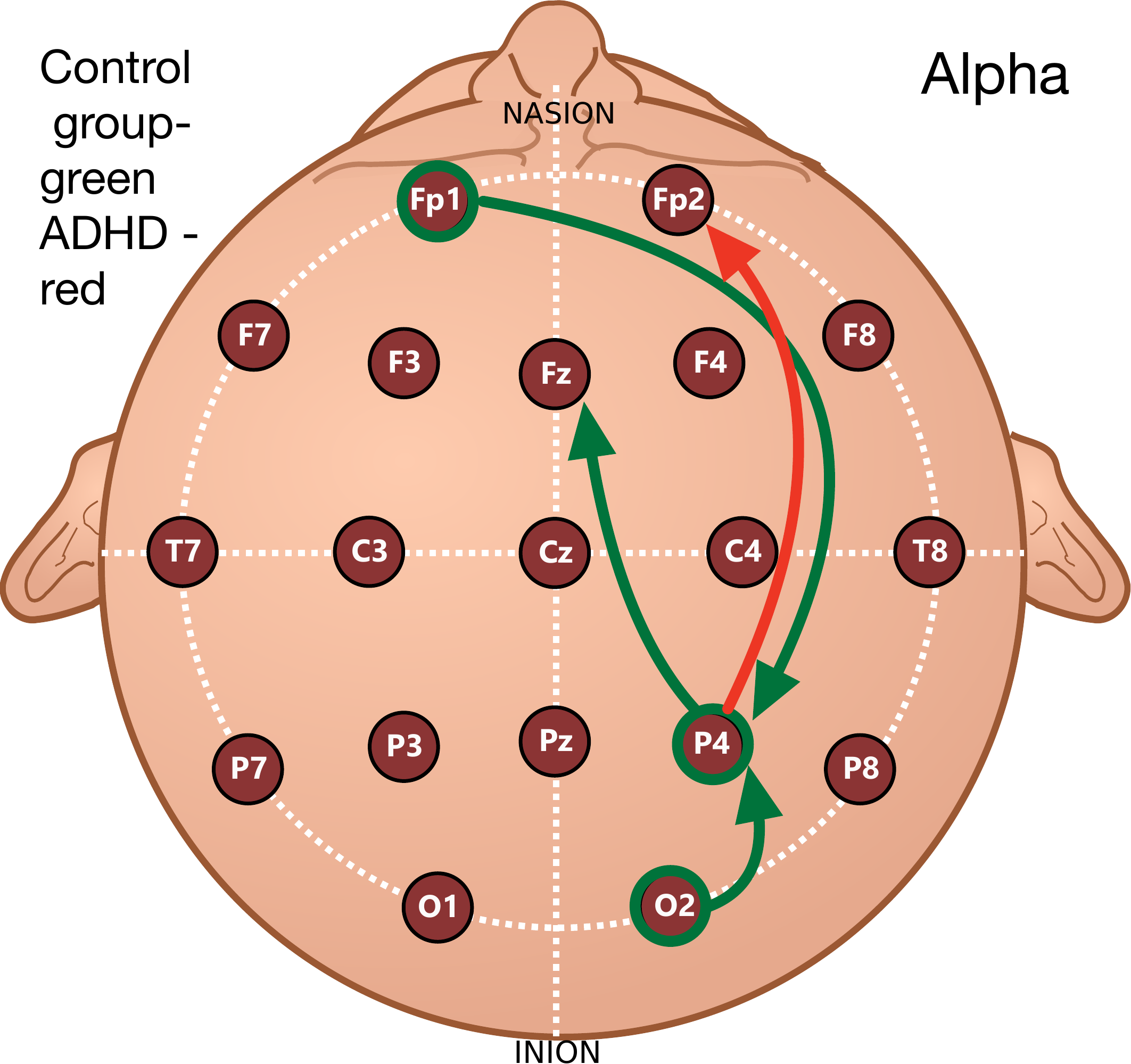}&
		\includegraphics[scale=0.14]{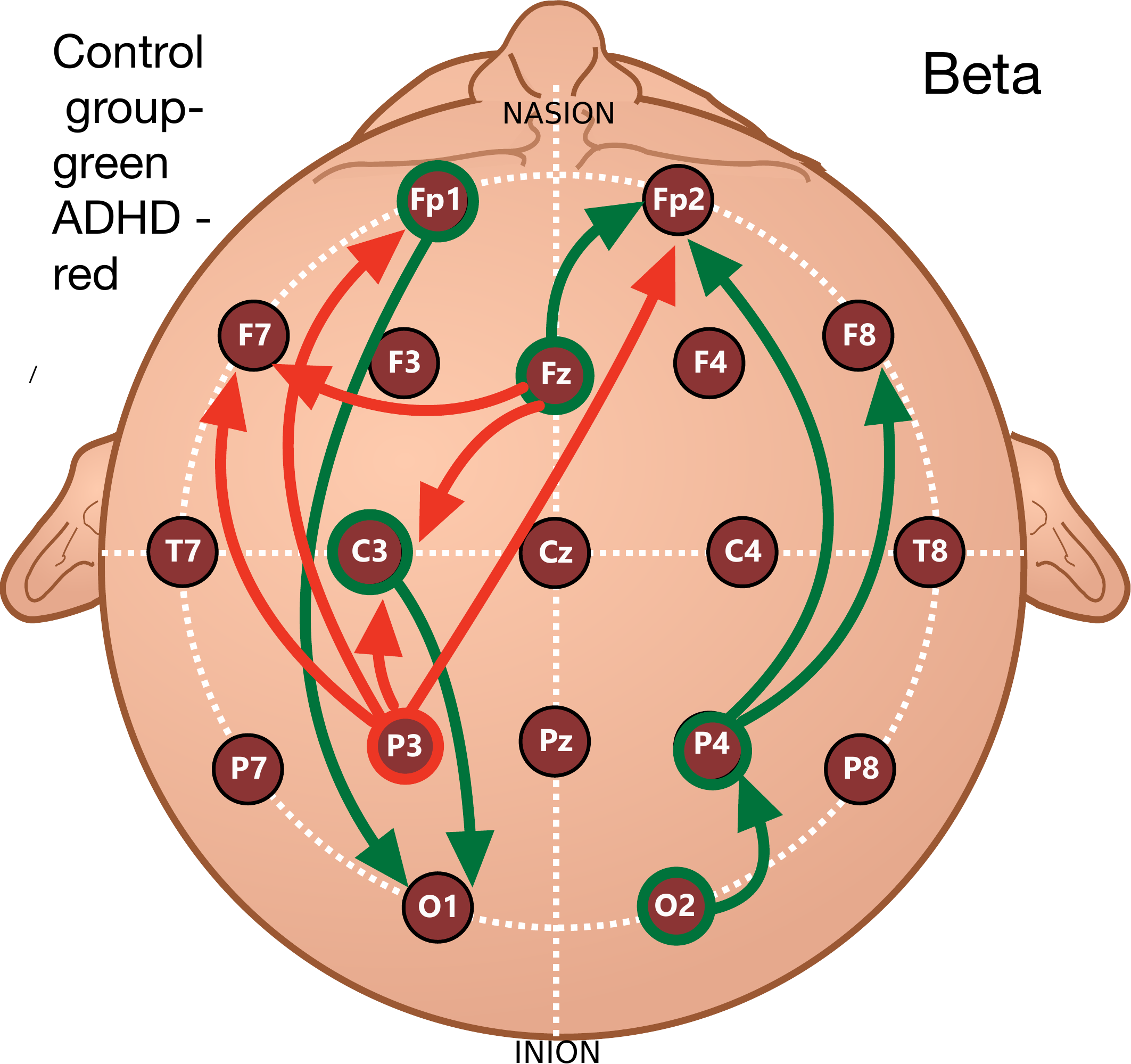}&
		\includegraphics[scale=0.14]{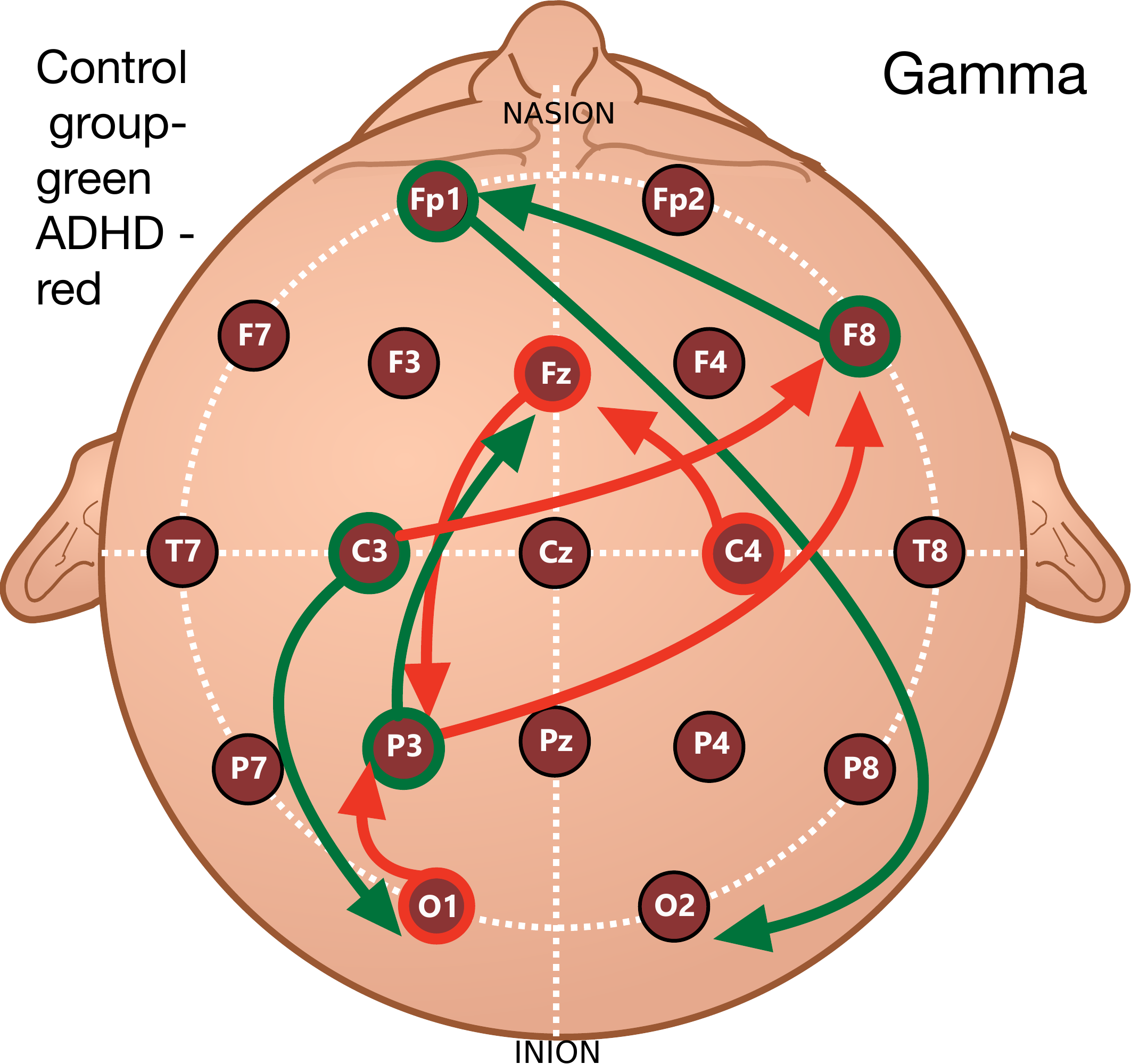}
	\end{tabular}
	\vspace{4mm}
	\caption{Part (a): Fixed connectivity parameters represent Granger causality in each group and frequency band. Arrows represent significant connections with p-value less than $10^{-6}$ and magnitude of coefficient is higher than 80\% quantile of empirical distribution of estimates. Arrow from $Ch_1$ to $Ch_2$ represents that $Y_{Ch_1}(t-1)$ component significantly improves predictability of $Y_{Ch_2}(t)$. Part (b): Differences in Fixed connectivity parameters. Green connections are presented in control group, but are not presented in ADHD, red shows presented in ADHD, but not presented in control group.}
	\label{fig:connectiv}
\end{figure}

\begin{figure}[!htp]
	\centering
    \title{Standard deviations of random effects and differences between groups}\par
     \vspace{4mm}
		\includegraphics[scale=0.33]{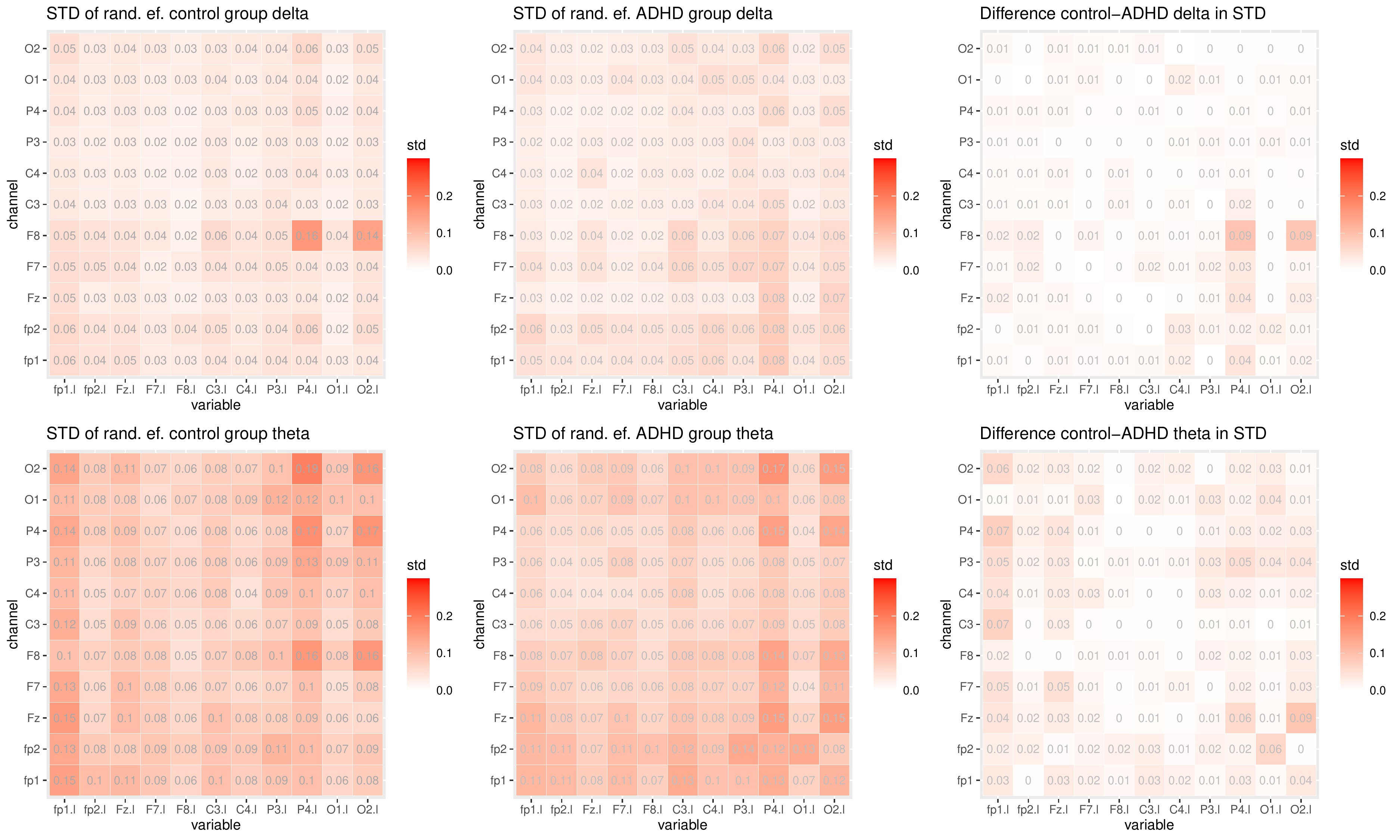}
		\includegraphics[scale=0.33]{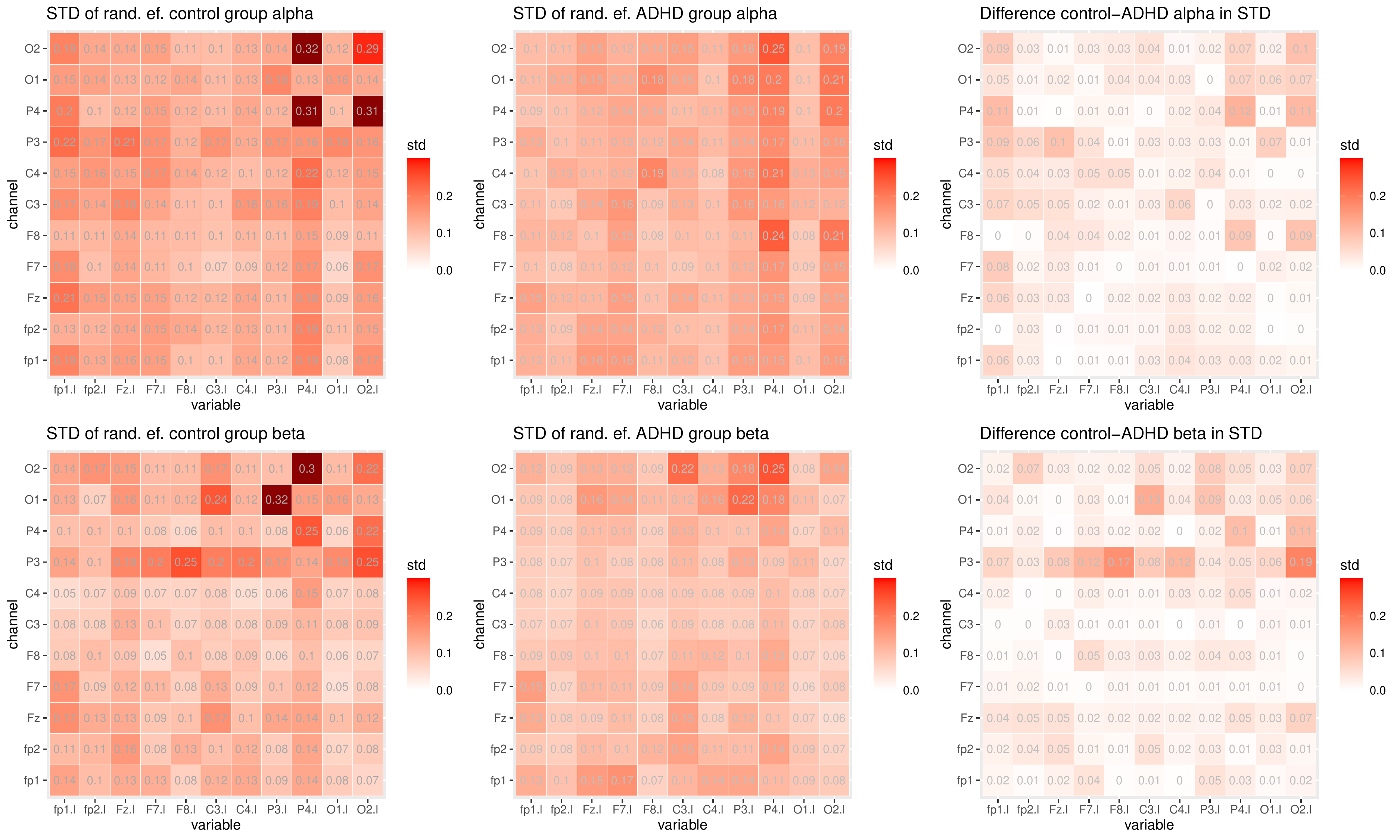}
		\includegraphics[scale=0.33]{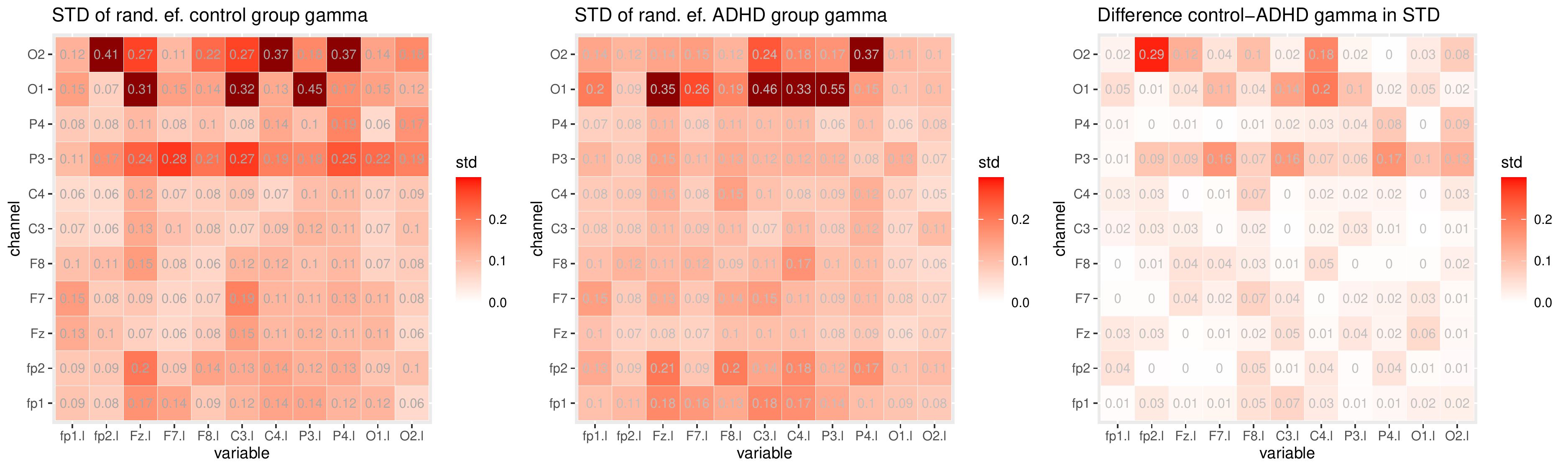}
	
	\vspace{4mm}

	\caption{Standard deviations of random effects obtained by ME-SpecVAR in five frequency bands. Third column represents magnitude of differences between control group and ADHD group. Darker color corresponds to higher difference.}
	\label{fig:different}
\end{figure}

Under the ME-SpecVAR model, the fixed effects are group specific connectivity matrices $\Phi^{band}_{1, 1}$ and $\Phi^{band}_{2, 1}$ in two groups and $b_1^{band}$, $b_2^{band}$ matrices of standard deviations of the random effects for each frequency $band = (\delta, \theta, \alpha, \beta,\gamma)$.
The estimation procedure is based on restricted maximum likelihood using lme4 package in R \citep{lme4}. We estimated the parameters using 11 channel data decomposed to common five frequency bands, showed in Figure \ref{fig:eeg}. 
As a result, we obtained group specific connectivity parameters for each frequency band via ME-SpecVAR and standard deviations of random effects. Significant connectivity parameters are illustrated alternatively as a graph of connections between brain regions in Figure \ref{fig:connectiv}, (a). Significance level was corrected to $p < 10^{-6}$ in order to avoid the problem of multiple comparisons using Bonferroni correction. In addition, we are interested in features that help to discriminate groups with each other. Such features are unique significant connections which are presented in one group but not presented in another in Figure \ref{fig:connectiv}, (b). Standard deviations of random effects in five frequency bands are illustrated as heatmaps and difference between control and ADHD group is also obtained in Figure \ref{fig:different}.

\subsection{Discussion of the results}
The model was proposed to capture group specific parameters and subject specific random deviations of the parameters from children with ADHD and healthy controls. Such results give us effective connectivity measure as well as level of homogeneity of connectivity in each group during visual attention task. After the investigation of best lag value and conducting simulation study we decided to proceed with ME-SpecVar (1) and cover the frequency related estimates. 

General patterns of connectivity and flow of information are seen between right anterior and left posterior regions in slow waves and from parietal region to frontal in fast frequency bands. The evident difference in general patterns of flow is in the absence of connectivity from prefrontal region to parietal-occipital in ADHD group compared to control group in Figure \ref{fig:connectiv}, (b). The connectivity in that regions is associated with conscious visual attention and control of selective and focused attention \citep{GOTTLIEB2010240}.
 Significant connectivity estimates are in general more presented in healthy children in all frequency bands in Figure \ref{fig:connectiv}, (b), which is consistent with findings about overall lack of connectivity in ADHD \citep{10.1117/1.NPh.7.1.015012}. 
ME-Spec-VAR provides us with standard deviations of random effects in each group for each effective connectivity measure, which is unique finding that was not accounted in previous studies. We revealed that most of the differences appear at P3 (parietal) channel and occipital O1 and O2 channels in beta and gamma bands in Figure \ref{fig:different}. Control group variability of random effects is higher in fast frequencies beta and gamma according to the results of our paper. Occipital O2 channel and parietal P4 channels influence to F8 channel is higher in control group at slow frequencies, that is shown in Figure \ref{fig:different} at delta, theta and alpha. The parietal and occipital channels differences in ADHD are also emphasized in previously mentioned articles. One possible reason of elevated variability in control group at beta and gamma bands could be task specific performance, in which ADHD group showed more consistent and decreased pattern of connectivity, whereas in healthy children the connectivity is higher and the variability in the group is elevated due to more diverse performance of higher-order cognitive functions.

 Additionally, there are few studies that  assess temporal variability of functional connectivity in ADHD \citep{ALBA20161321, Barttfeld2014}. The results revealed enhanced variability in ADHD group compared to control group. However, in our case we do not reveal temporal variability, we find the variability of subject specific parameters in the group. The difference between our approach and the studies above is also that they measured not effective but functional connectivity using synchronization likelihood method and averaged coherence. Another essential difference is that the experimental paradigm is resting state, however we use dataset based on attention task. This makes the results revealing different kinds of variability in our study and aforementioned work.

%% file: 4_conclusion.tex
\section{Conclusion}\label{chap:conclusion}

\noindent The main contribution of this paper is the new model ME-SpecVAR, which is a generalization of mixed effects vector autoregressive model for applications in different frequency oscillations. It provides additional results of effective connectivity in ADHD children. The advantage of the ME-SpecVAR model is that the model provides estimates of fixed and random effects in terms of oscillatory activities. Fixed effects are group specific fixed connectivity parameters at each frequency band. They can be used for testing Granger causality of EEG channels activity on each other. The random effects are subject specific variations of fixed parameters in each group and each oscillatory activity, that show the level of variability and homogeneity of the parameters in each group. The ME-SpecVAR model produced new and interesting results. The current approach allows to conduct one stage estimation of fixed and random parameters in different frequency bands and to test for significant parameters at the same time, namely find Granger Causality. Results obtained from different frequency components give new insights into ADHD and healthy children brain effective connectivity. The overall decreased connectivity at all oscillations and decreased variability of subject specific parameters is inherent to ADHD group based on Granger causality tests. This new perspective allows to see whether the connection between brain regions is mostly group characterizing or it includes the natural variability of participants between each other. Higher variability of parameters in control group can refer to enhanced diversity of connections as a normal condition, whereas ADHD parameters have less between subject variability, that can be attributed to disorder specific feature. The observed differences in variability could mean that similar patterns of connectivity are more inherent to children diagnosed with ADHD.  Results revealed brain effective connectivity structures from both groups, that share flow of information between left anterior and right posterior regions in slow waves and from parietal region to frontal in fast frequencies. Distinctions between groups are connections from prefrontal region to parietal-occipital in control group that ADHD group lacks. The Parietal-Occipital  brain region connection is associated with control of selective and focused visual attention, and most variability and connectivity is referred to channels of that region.  There are some difficulties during comparison of the results with previous studies. Different experimental procedures as resting state with closed eyes, open eyes and cognitive tasks influence on the results \citep{Barttfeld2014}. The measures of functional and effective connectivity are different, that also causes diverse results. The models and methodologies are different as well in the described works. All of the aspects make the comparison of results problematic and challenging for the scientists to infer the patterns. Nevertheless, many findings of our research are consistent with previous studies and our study complements the existing picture of ADHD research.  Currently, our method did not incorporate the higher lag orders and whole brain electrodes due to computational difficulties of restricted maximum likelihood estimation of fixed and random effects. In future work, it is essential to optimize the procedure of estimation in order to be able to add more parameters into the model.